\documentclass[journal]{IEEEtran}
\usepackage{hhline}

 \usepackage{cite}
\ifCLASSINFOpdf
\else
   \usepackage[dvips]{graphicx}
   \graphicspath{{./}}
   \DeclareGraphicsExtensions{.eps}
\fi
\usepackage[cmex10]{amsmath}
\usepackage{url}
\usepackage{amssymb, amsthm}
\usepackage{psfrag}

\usepackage{color}

\newcommand{\paino}{\sl} 
\newcommand{\wei}{u} 
\newcommand{\pdim}{p}
\newcommand{\ndim}{n}

\renewcommand{\dim}{p} 
\newcommand{\covm}{\bo R} %\bom{\mathcal C}} % Covariance matrix 
\newcommand{\V}{\bo V}     % SHAPE MATRIX 

\newcommand{\A}{\bo A}
\renewcommand{\P}{\bo P}
 \newcommand{\B}{\bo B}
 \newcommand{\K}{\bo K}
 
 \renewcommand{\S}{\hat{\bo R}}

\newcommand{\I}{\bo I} 
\newcommand{\y}{\bo y} 
 
\renewcommand{\u}{\bo u}

\newcommand{\z}{\bo z}
\newcommand{\0}{\bo 0}
\newcommand{\p}{\bo p}
\renewcommand{\c}{\bo c}

\newcommand{\Lam}{\bom \Lambda} 
\newcommand{\Lamm}{\Lambda} 

\newcommand{\be}{\beta}
\newcommand{\al}{\alpha}  
\newcommand{\sigw}{\wei}   % Weight, sometimes denote u(t) in M-es. eq : M = ave{ u(x' M^{-1} x) x x' } 
\newcommand{\gam}{\gamma}  
\newcommand{\lam}{\lambda}  
\newcommand{\tha}{\bom \theta} 
\newcommand{\bV}{\bo V}
\newcommand{\hS}{\widehat{\bom \Sigma}}
\newcommand{\eh}{\frac{1}{2}}
\newcommand{\x}{\bo x}

\newcommand{\beq}{\begin{equation}}
\newcommand{\eeq}{\end{equation}}
\newcommand{\bmat}{\begin{pmatrix}}
\newcommand{\emat}{\end{pmatrix}}
 \newcommand{\beqa}{\begin{eqnarray}}
\newcommand{\eeqa}{\end{eqnarray}}

\newcommand{\CE}{\mathrm{CE}}
\newcommand{\CN}{\C \mathcal N}

\newcommand{\M}{\bom \Sigma}
\newcommand{\Sig}{\bom \Sigma}

\newcommand{\hop}{\mathrm{H}}              % Hermitian transpose 
\newcommand{\R}{\mathbb{R}}      %  Field of real numbers
\newcommand{\C}{\mathbb{C}}      %  Field of real numbers
\newcommand{\tr}{\mathrm{Tr}}     % Matrix trace 
\newcommand{\Tr}{\mathrm{Tr}}     % Matrix trace 
\newcommand{\PDH}{\mathcal{H}}   % Class of PDH matrices
\newcommand{\hypo}{\mathop{\gtrless}^{H_1}_{H_0}}

\newcommand{\bom}[1]{\boldsymbol{#1}}    % boldface math (for greek letters)
\newcommand{\bo}[1]{\mathbf{#1}}              % boldface math 
\newcommand{\diag}{\mathrm{diag}}             % diagonal matrix 
\newcommand{\dist}{\mathcal D}  
\newcommand{\Mn}{\M_0}  
\newcommand{\Mal}{\M_{\al}}

\newcommand{\E}{\mathbb{E}}

\newcommand{\mM}{\mathcal{M}}
\newcommand{\mV}{\mathcal{V}}

\begin{document}
\title{Regularized  $M$-estimators of scatter matrix} 

%CFAR detection under complex elliptically symmetric noise

%Circularity detection under complex elliptically symmetric distributions} 
%distributions - Part II: complex normal distribution and its extensions}


\author{Esa~Ollila,~\IEEEmembership{Member,~IEEE,}~and~David~E.~Tyler% <-this % stops a space
%\thanks{Copyright (c) 2012 IEEE. Personal use of this material is permitted.
%However, permission to use this material for any other purposes must be
%obtained from the IEEE by sending a request to pubs-permissions@ieee.org.}% <-this % stops a space
\thanks{E. Ollila is 
with the Department
of Signal Processing and Acoustics, Aalto University, Espoo,  
P.O. Box 13000, FIN-00076  Aalto; 
e-mail: esa.ollila@aalto.fi (see http://signal.hut.fi/\~{}esollila). D.E.~Tyler is with 
the Department of Statistics \& Biostatistics, Rutgers -- The State University of New Jersey, Piscataway NJ 08854, USA; e-mail:  dtyler@rci.rutgers.edu.}% <-this % stops a space
\thanks{Manuscript received May 11th, 2014}}
%; revised XXX, 201X.


  \markboth{ Submitted to IEEE TRANSACTIONS ON SIGNAL PROCESSING}{E. Ollila and D. E. Tyler:  Regularized $M$-estimators of scatter matrix}  
%in radar and senr



\newtheorem{definition}{Definition}
\newtheorem{lemma}{Lemma}
\newtheorem{theorem}{Theorem}
\newtheorem{corollary}{Corollary}
\newtheorem{condition}{Condition}

\maketitle

\begin{abstract} 
In this paper, a general class of regularized $M$-estimators of scatter matrix are proposed which are suitable also for low or insufficient sample support (small $\ndim$ and large $\pdim$) problems. 
The considered  class constitutes a natural generalization of $M$-estimators of scatter matrix (Maronna, 1976) and are defined  as a solution to a penalized $M$-estimation cost function that depend on a pair $(\al,\be)$ of regularization parameters.  We  derive general conditions for uniqueness of the solution using concept of geodesic convexity. Since these conditions do not include Tyler's $M$-estimator, necessary and sufficient conditions for uniqueness of the penalized Tyler's cost function are established separately.  
For the regularized Tyler's $M$-estimator, we also derive a simple, closed form and data dependent solution for choosing the regularization parameter based on shape matrix matching in the mean squared sense.  
An iterative algorithm  that converges to the solution of the regularized $M$-estimating equation is also provided.   
Finally, some simulations studies  illustrate the  improved accuracy of the proposed regularized $M$-estimators of scatter compared to their non-regularized counterparts in low sample support problems. An example of radar detection using normalized matched filter (NMF) illustrate that an adaptive NMF detector based on  regularized $M$-estimators are able to maintain accurately the preset CFAR level and at at the same time provide similar probability of detection as the (theoretical) NMF detector. \end{abstract}


% \fi
\IEEEpeerreviewmaketitle


\section{Introduction}

\IEEEPARstart{M}{any}  data mining and classic multivariate analysis techniques require an estimate of the covariance matrix or some nonlinear function of it, e.g., 
the inverse covariance matrix or its eigenvalues/eigenvectors. Given an  i.i.d.  sample $\z_1,\ldots,\z_n \in \C^{\pdim}$ from a centered, 
i.e.,~$\E[\z] = \0$, (unspecified) $\pdim$-variate distribution $\z \sim F$, the {\paino sample covariance matrix} (SCM)
$\S=\frac{1}{\ndim}\sum_{i=1}^\ndim \z_i \z_i^\hop  \in \C^{\pdim \times \pdim}$ 
is the most commonly used estimator of the unknown covariance matrix $\covm=\E[\z \z^\hop]$. However, in  high-dimensional (HD) problems, there are many 
cases that the SCM simply can not be computed, is completely corrupted, or is inaccurate. For example, low sample support (LSS) (i.e.,  $\pdim$ is of 
the same magnitude as $\ndim$)  is a commonly occurring problem in diverse HD data analysis problems such as %micro-array analysis of gene expression, 
chemometrics and medical imaging.  
In the case of {\paino insufficient sample support (ISS)}, i.e., $\pdim> \ndim$,  the inverse of the SCM 
can not be computed.  Thus, for example, 
% PCA, which is a common pre-processing
%and standardization step in many multivariate procedures, cannot be performed due to rank deficient SCM. 
%Also 
classic beamforming techniques such as MVDR beamforming 
or the adaptive normalized matched filter cannot be realized  
since they require an estimate of the inverse covariance matrix. 

Robust estimation is also a key property  in HD data analysis problems. Partly because outliers are more difficult to glean from HD data sets by conventional 
techniques, but also due to an increase of impulsive measurement environments and outliers in practical sensing systems.  
The SCM is well-known to be vulnerable to outliers and to be a highly inefficient estimator when the samples are drawn from a heavy-tailed non-Gaussian distribution. HD data poses additional problems and difficulties since most robust estimators  such as $M$-estimators of scatter matrix \cite{maronna:1976} can not be computed in ISS scenarios,
or are equivalent to the SCM \cite{tyler:2010}. 

In this paper, we address this issue and propose a general class of regularized $M$-estimators of scatter matrix. 
This class provides  practical and  actionable estimators of the covariance (scatter) matrix even in the problematic ISS case. The proposed class constitutes a natural generalization of $M$-estimators of scatter \cite{maronna:1976} and their complex-valued generalizations \cite{ollila_koivunen:2003a,ollila_etal:2012}, and are defined  as a solution to a penalized $M$-estimation cost function that includes a pair $(\al,\be)$ of fixed regularization parameters.   We  derive a general conditions for uniqueness of the  solution using theory of geodesic convexity which has been previously 
utilized in \cite{wiesel:2012,zhang_etal:2013} in studying the uniqueness of the non-regularized $M$-estimators of scatter whereas  \cite{wiesel:2012b} focused on the regularized Tyler's $M$-estimator of scatter matrix  
using a particular scale invariant geodesically convex penalty function.  Our class include as special case,  the cost function for  $\pdim$-variate complex normal samples, for which the unique solution of the penalized cost function is easily found to   
\beq \label{eq:GLC} 
\S_{\al,\be} = \be \S + \al \I ,
\eeq  
which in \cite{du_etal:2010}, was called as the general linear combination (GLC) estimator. It should be noted however that  
in \cite{du_etal:2010}, $\S_{\al,\be}$ was not proposed as a minimizer to any optimization problem.

Our general conditions do not apply to the cost function corresponding to Tyler's  \cite{tyler:1987} $M$-estimator and hence this estimator is treated seperately, with necessary and sufficient conditions being established to ensure the uniqueness of solution for the penalized Tyler's cost function. Regularized versions of Tyler's $M$-estimator have also been recently studied in \cite{pascal_etal:2013} for the case $\beta = 1-\alpha$ and under more strict conditions on the sample, and also in \cite{chen_etal:2011},  but not in the
context as a solution to a penalized $M$-estimation cost function.  Estimation of the regularization parameters 
using the expected likelihood approach was proposed in \cite{abramovich_ besson:2013,besson_abramovich:2013} for  the regularized Tyler's $M$-estimator of \cite{chen_etal:2011},  
whereas \cite{couillet_mckay:2014} based their analysis on random matrix theory (both $n$ and $p$ are large). 
  For the regularized Tyler's $M$-estimator, we also derive a simple, closed form and data dependent solution to compute the regularization parameter $\al$ based on shape matrix matching in the mean squared sense.  %, which can be viewed as a diagonally-loaded constrained Tyler's $M$-estimator.  
We illustrate the usefulness of the regularized $M$-estimators of scatter in radar detection application  using normalized matched filter. Finally, we note that although our derivations are for complex-valued case, they generalize in an straightforward manner to real-valued case as well.  

The paper is organized as follows. Section~\ref{sec:prelim} reviews complex elliptically symmetric (CES) distributions and the maximum likelihood (ML) and $M$-estimators of the scatter matrix parameters of the CES distributions \cite{ollila_etal:2012}. Section~\ref{sec:regM} then introduces the  penalized $M$-estimation cost function. The stationary points are shown to be solutions to shrinkage type 
$M$-estimation equations.  Interpretation of regularization parameters are discussed and specific examples of regularized $M$-estimators are given. In Section~\ref{sec:uniq}, general  conditions  are presented to ensure the uniqueness of solution, with the proof of uniqueness being based on the concept of geodesic convexity. The regularized Tyler's $M$-estimator is then considered in Section~\ref{sec:tylM} and numerical examples are given in Section~\ref{sec:simul}.  Some of the proofs are reserved for the Appendix.

{\it Notations:} Let $\PDH(\pdim)$  denote the class positive definite Hermitian (PDH) $\dim\times \dim$ matrices, $| \bo A |$ the determinant of a square matrix $\bo A$. Furthermore, $\| \cdot \|$ (resp. $\| \cdot \|_1$) denotes the $\ell_2$-norm (resp. $\ell_1$-norm) 
%(or {\paino Frobenius matrix norm}),  
defined as $\| \A \|^2 = \Tr(\A^\hop \A)=\sum_i \sum_j |a_{ij}|^2 $ 
(resp. $\| \A \|_1 =\sum_i \sum_j |a_{ij}|$)  for any  $m \times n$ matrix $\A$.

\section{Preliminaries} \label{sec:prelim} 

\subsection{Elliptical distributions} 

A continuous symmetric random vector (r.v.)  $\z \in \C^\pdim$ has a  centered {\paino complex elliptically symmetric (CES) distribution} \cite{ollila_etal:2012} if its p.d.f. is of the form:   
\[
f(\z) = C_{\pdim,g}  |\M|^{-1} g\big( \z^\hop \M^{-1} \z \big),  
\]
where $\M \in  \PDH(\pdim)$  is the unknown parameter, called the {\paino scatter matrix},   $g:\R_0^+ \to \R^+$ is a  fixed function  called the {\paino density generator} 
and $C_{\dim,g}>0$ is a normalizing constant ensuring that $f(\z)$ integrates to one. 
We denote this case by $\z \sim \CE_\dim(\bo 0, \M,g)$. 
If the covariance matrix $\covm= \E[\z \z^\hop]$ of $\z$ exists, then  
 \[
 \covm= c \cdot \M \quad (\mbox{for some } c >0). 
 \] 
 For example, when $g(t)=\exp(-t)$, one obtains the $\pdim$-variate complex normal (CN) distribution, denoted
$\z \sim \mathcal \CN_\pdim(\bo 0, \M)$; In this case, $\covm=\M$.  
For a detailed account on properties of CES distributions, we refer the reader to  \cite{ollila_etal:2012}. 
Let  $\z_1,\ldots,\z_n$ denote an i.i.d. random  
sample from an unspecified $p$-variate CES distribution as stated above. 

The maximum likelihood estimator (MLE) of scatter matrix, denoted $\hat \M$, minimizes the negative log-likelihood function (divided by $n$)
\beq \label{eq:cost}
\mathcal L(\M)=  \frac{1}{\ndim} \sum_{i=1}^\ndim \rho( \z^\hop_i \M^{-1} \z_i ) -   \ln |\M^{-1}| 
\eeq 
where  $\rho(t)= -\ln g(t)$. More appropriate notation would be $\mathcal L_n(\M | \rho)$ to emphasize  the dependence on $\rho$ and the sample.  
Critical points are then solutions to the estimating equation
\beq \label{eq:esteq}
\hat \M=\frac{1}{\ndim} \sum_{i=1}^\ndim \wei(\z_i^\hop \hat \M^{-1} \z_i) \z_i \z_i^\hop
\eeq
where $\wei= \rho'=-g'/g$. 

\subsection{$M$-estimators of scatter}

 {\paino $M$-estimators} of scatter are generalizations of the ML-estimators of the scatter matrix of an elliptical distribution. 
They can be defined by allowing a general $\rho$ functions in \eqref{eq:cost}, not necessarily related to any elliptical density $g$, in
which case we refer to \eqref{eq:cost} as a general cost function. The function $\rho$ is usually chosen so that the corresponding
weight function $\wei=\rho'$ is non-negative, continuous and non-increasing. Equation \eqref{eq:esteq} is then referred to as an $M$-estimating equation.
Some examples of $M$- and ML-estimators are given below. 

{\it SCM (the Gaussian MLE)}. In the Gaussian case,  $\rho(t)=t$ and $\wei(t) =\rho'(t)\equiv 1$, so eq. \eqref{eq:cost} becomes 
\[
\mathcal L(\M) = \tr( \S \M^{-1}) - \ln | \M^{-1} |
\]
where $\S$ denotes the SCM. The (well-known) {\it unique} minimizer (assuming $\ndim \geq \pdim$) of this function is the sample covariance matrix, i.e., $\hat \M = \S$.

{\paino Complex Tyler's \cite{tyler:1987} $M$-estimator}  is based on the functions 
\[
\rho(t)= \pdim \ln t \quad  \mbox{and} \quad \wei(t)= \rho'(t)=\frac{\pdim}{t}. 
\]
Note that this $\rho$-function is {\it not} related to any elliptical density and the optimization problem \eqref{eq:cost} is now non-convex. 
Nevertheless, the estimator is actionable: a unique solution (up to a scale) exists under mild conditions and the global solution can be 
computed via simple fixed-point iterations; see \cite{tyler:1987,pascal_etal:2008,ollila_etal:2012}. It should be noted that for Tyler's $M$-estimator, the summations in both \eqref{eq:cost} and \eqref{eq:esteq} are taken only over $\z_i \ne \0$.
In the radar community, Tyler's $M$-estimator is often referred to as a fixed-point estimator, and it is known to admit numerous ML-interpretations 
as shown in  \cite{tyler:1987b,kent:1997,gini_greco:2002, conte_etal:2002, ollila_tyler:2012} in the real and complex cases.

{\paino Complex Huber's $M$-estimator} is based on a weight function of the form \cite{ollila_koivunen:2003c} 
\[
\wei(t) = \begin{cases}  1/b,  
&  \ \mbox{for} \ t \leq c^2 \\ c^2/(t b),  & \ \mbox{for} \ t > c^2 \end{cases} 
\]
where $c$ is a tuning constant defined such that $q=F_{\chi^2_{2\dim}}(2c^2)$ for a chosen $q$ ($0<q\leq 1$), where $F_{\chi^2_{2\dim}}(\cdot)$ denotes the c.d.f. of the chi-squared 
distribution with $2\dim$ degrees of freedom. The scaling factor 
$b$ is usually chosen so that the resulting $M$-estimator is consistent to the covariance matrix for Gaussian data, 
namely $b = F_{\chi^2_{2(\dim+1)}}(2c^2) + c^2(1-q)/\dim$.  If $q \to1$, then Huber's estimator approaches the SCM, and if 
 $q \to 0$, then the estimator approaches Tyler's $M$-estimator.

\section{Regularized $M$-estimators of scatter matrix} \label{sec:regM}

To stabilize the optimization problem an additive penalty term $\al \cdot \mathcal P(\M)$ can be introduced to  the cost function \eqref{eq:cost}, 
where $\al \geq 0$ denotes a fixed regularization parameter. A popular focus in the literature has been to enforce
sparsity on the {\paino precision matrix} $\K = \M^{-1}$ by using $\ell_1$-penalty function 
\beq \label{eq:l1pen} 
\mathcal P_{\ell_1}(\M) = \| \M^{-1} \|_1 
\eeq  
as is done in the real-valued case in \cite{witten_tibshirani:2009,friedman_etal:2008}. The use of the  
$\ell_1$-penalty, though, to help enforce a sparse precision matrix is dependent on the cost function \eqref{eq:cost} being convex in
$\M^{-1}$, which holds whenever $\rho(t)$ itself is convex. However, robust $M$-estimates of scatter typically have 
decreasing weight functions $\wei(t)$ and hence concave $\rho$-functions.
  
In this paper, we take a different approach and focus on a penalty function of the form 
\begin{align*} 
\mathcal{P}^*(\M) & = \| \M^{-1/2} \|^2  %\\ %=  \mbox{ sum of squares of the elements } \\ 
=\tr( \M^{-1}). 
\end{align*} 
Notice that 
\[
\tr(\M^{-1})=\sum_{j=1}^\pdim \frac{1}{\lam_j(\M)},
\]
 where $\lam_j(\M)$'s denote the ordered eigenvalues of $\M$. Thus the penalty term restricts  $\frac{1}{\lam_j(\M)}$ from growing without bound;  this is necessary  in the ill-conditioned ISS case ($\ndim < \pdim$).  In addition to the additive penalty term $\al \mathcal P(\M)$, we impose a weight $\be$ on the cost term $ \sum_{i=1}^\ndim \rho(\z_i^\hop \M^{-1} \z_i)$, 
and thus our  {\paino penalized cost function} is of the form  
\beq  \label{eq:gpenfun} 
\mathcal L_{\al,\be} (\M) = \frac{\be}{n}  \sum_{i=1}^\ndim \rho( \z^\hop_i \M^{-1} \z_i ) -  \ln |\M^{-1}| + \al \mathcal P(\M),
\eeq
where  $\be >0, \al\geq 0$ form the pair of (fixed) regularization parameters.  For the case $\mathcal{P}(\M) = \mathcal{P}^*(\M)$ this becomes 
\beq
\mathcal L^*_{\al,\be} (\M) =\frac{\be}{n}  \sum_{i=1}^\ndim \rho( \z^\hop_i \M^{-1} \z_i ) -  \ln |\M^{-1}| + \al \tr(\M^{-1}) 
 \label{eq:penfun} 
\eeq

%Alternatively, one may write \eqref{eq:penfun} in the form
%\[
%\mathcal L_{\tilde \al,\be} (\M) = \frac{1}{n}  \sum_{i=1}^\ndim \rho( \z^\hop_i \M^{-1} \z_i ) -  \ln |\M^{-1}|^{1/\be} + \tilde \al \tr(\M^{-1}) 
%\]
%where $\tilde \al= \al/\be$. 
As will be illustrated below the parameter $\al$ can be best described as {\paino ridge (or spherizing)} parameter,
and the parameter $\be$ can be best described as a {\paino robust tuning} parameter.

Let $\hat \M$ denote the minimizer of $\mathcal L^*_{\al,\be} (\M)$.  %$\hat \M=\arg \min \mathcal L_{\al,\be}(\M)$. 
The solution $\hat \M$ naturally depends on 
$(\al,\be)$ but these are not made explicit for notational convenience.   
It is easy to verify using matrix differential rules that a critical point of the penalized cost function $\eqref{eq:penfun}$ is a solution to
\beq \label{eq:penMest} 
\hat  \M = \frac{\be}{n} \sum_{i=1}^\ndim \wei(\z_i^\hop \hat \M^{-1} \z_i) \z_i \z_i^\hop + \al \I
\eeq
which is  weighted and diagonally loaded form of the classic $M$-estimating equation obtained with $(\al,\be)=(0,1)$.  
Expressing the regularized $M$-estimating equation in the form 
\[
\I =  \frac{\be}{n} \sum_{i=1}^\ndim \wei(\z_i^\hop \hat \M^{-1} \z_i)\hat{ \M}^{-1} \z_i  \z_i^\hop + \al \hat \M^{-1},
\] 
and then taking the  trace shows that the solution $\hat \M$ must satisfy
\beq \label{eq:cond} 
\al \tr(\hat \M^{-1})= \pdim - \be \cdot  \Big \{ \frac{1}{\ndim} \sum_{i=1}^\ndim  \psi( \z_i^\hop \hat \M^{-1} \z_i)  \Big \}
\eeq
where $\psi(t)=t \sigw(t)$. 

{\bf Algorithm.} The regularized $M$-estimating equation \eqref{eq:penMest} gives rise to the following fixed point algorithm. 
Given any  initial value $\M_0 \in \PDH(\pdim)$,  iterate 
\beq \label{eq:algor}
\hat \M_{k+1} =  \frac{\be}{n} \sum_{i=1}^\ndim \wei(\z_i^\hop \hat \M^{-1}_{k} \z_i) \z_i \z_i^\hop + \al \I
\eeq 
until convergence. 
The algorithm converges to a solution of \eqref{eq:penMest} given any initial value
$\hat\M_0$. The proof of convergence is analogous to the convergent
proof for the non-regularized $M$-estimators given in \cite{kent_tyler:1991} and is given in the Appendix. 
For convergence of the algorithm we need to  assume that $\rho(t)$ is continuously differentiable and satisfies Condition 1 (stated below in Section~\ref{sec:uniq}) 
and that the $M$-estimating equation \eqref{eq:penMest} has a unique solution $\hS$. 
Conditions for uniqueness are given in Theorems 1, 2 and 3.

%Note that if the algorithm converges then it must converge to a solution of \eqref{eq:penMest}. 

The interpretation of $\be$ as a robust tuning parameter follows by expressing \eqref{eq:penMest} in the form
\[
\hat  \M_\be = \frac{1}{n} \sum_{i=1}^\ndim \wei_\be(\z_i^\hop \hat \M_\be^{-1} \z_i) \z_i \z_i^\hop + \tilde \al \I,
\]
where $ \hat \M_\be = \hat \M/\be$, $\wei_\be(t) = \wei(t/\be)$ and $\tilde \al = \al/\be$. In particular, note that if $\wei(t)$ 
corresponds to Huber's weight function with a tuning constant $c$, then $\wei_\be(t)$ corresponds to Huber's weight function with 
a tuning constant of $\tilde c = c~\be^{1/2}$. A more detailed discussion on tuning weight functions can be found 
in \cite{kent_tyler:1996}. 
For the two extreme cases $c \rightarrow \infty$ and $c \rightarrow 0$, which correspond to a regularized SCM and Tyler's
M-estimate respectively, the role of $\be$ is more subtle. We consider these special cases below.

{\paino GLC estimator}. In the Gaussian case $\rho(t)=t$, the penalized cost function \eqref{eq:penfun} simplifies to the form 
\[
\mathcal L^*_{\al,\be} (\M) =   \tr \big\{ (\be \S + \al \I) \M^{-1} \big\} - \ln | \M^{-1} |
\]
where $\S = \frac{1}{\ndim} \sum_{i=1}^\ndim \z_i \z_i^\hop$ denotes the SCM. 
The unique minimizer $\hat \M$ of the function above is easily shown to the GLC estimator \eqref{eq:GLC}, i.e.,  $\hat \M=\S_{\al,\be}$. 
For $\be = 1$, the solution is the diagonally loaded SCM, $\S_{\al} = \S + \al \I$. The interpretation of the GLC estimator as a solution to an optimization 
problem \eqref{eq:penfun} %given here is new and so 
differs from the motivation for the GLS estimator given in \cite{du_etal:2010}.  
%This result illustrates that diagonally loaded SCM or GLC estimator 
%have often used in array/radar signal processing for obtaining an estimate of the inverse covariance matrix in LSS or HD-LSS scenarios. For
 Note that the eigenvalues of $\S_{\al,\be}$ are 
$\hat \lam_i = \be \hat \lam_{\S,i} + \al$,
where $\hat \lam_{\S,i}$, $i=1,\ldots,\pdim$ denote the eigenvalues of $\S$. Thus  $\al$ can  be  viewed as a ridge parameter as it provides a ridge down the diagonal and guarantees a non-singular solution.  
It can be also described as  a {\paino spherizing parameter} since the larger the $\al$, the more "spherical" is the solution (i.e., as $\al$ gets larger, $\hat \M$ is shrinked towards  a scaled identity matrix $ \alpha \I$ ).  

{\paino Regularized Tyler's $M$-estimator} uses the weight function $\sigw(t)=\pdim/t$ and hence corresponds to 
a solution to  
 \beq \label{eq:Tyl}
\hat  \M = \frac{\pdim\be}{n_*} \sum_{i=1,\z_i \neq \bo 0}^\ndim \frac{\z_i \z_i^\hop}{\z_i^\hop \hat \M^{-1} \z_i} + \al \I,
\eeq
where $n_{*} = \#\{\z_i \ne 0 ; i=1,\ldots,\ndim \}$.
Condition \eqref{eq:cond} implies $\tr(\hat \M^{-1})=\pdim(1-\be)/\al$ and hence the
choice $\be=1$ is excluded. If we choose $\be=1-\al$ above, then the estimator  
$\hat \M$ satisfies the constraint $\tr(\hat \M^{-1})=\pdim$. Hereafter, when using this estimator, we assume without loss of
generality that $n_* = n$. This case $\be=1-\al$ has been previously studied in \cite{pascal_etal:2013}. 

\section{Uniqueness and Geodesic convexity} \label{sec:uniq}

In this section, we show under general conditions that there exists a unique minimizer to the penalized likelihood or cost function given by \eqref{eq:penfun}. 
Hereafter, it is assumed that the function $\rho(t)$ satisfies the following condition.

\begin{condition} \label{cond:rho}
The function $\rho(t)$ is nondecreasing and continuous for $0 < x < \infty$. Also, $r(x) = \rho(e^x)$ is convex in $-\infty < x < \infty$
\end{condition}

Note that if the function $\rho(t)$ in differentiable, then the above condition holds if and only if the weight function $\wei(t) \ge 0$
and $\psi(t) = t \wei(t)$ is nondecreasing. It readily follows that Huber's and Tyler's $M$-estimators as well as Gaussian MLE satisfies
Condition~\ref{cond:rho}.

The concept of geodesic convexity for functions of PDH matrices plays a key role in our proof of uniqueness. This concept has been previously 
utilized in \cite{wiesel:2012,zhang_etal:2013} in studying the uniqueness of the non-regularized $M$-estimates of scatter and in \cite{wiesel:2012b} in the case of regularized Tyler's cost function. 
A review of geodesic convexity  for positive definite matrices can be found in the aforementioned papers as well as in \cite{sra_hosseini:2013}, wherein further references can be found.  
We briefly review here some important results.

Rather than treating the class $\PDH(\pdim)$ as a convex cone in $\C^{\pdim}$ and using notions from complex Euclidean geometry, one can treat $\PDH(\pdim)$
as a differentiable Riemannian manifold with the geodesic path from $\M_0 \in \PDH(\pdim)$ to $\M_1 \in \PDH(\pdim)$ being
\beq \label{eq:geopath}
\M_t = \M_0^{1/2}\left(\M_0^{-1/2}\M_1\M_0^{-1/2}\right)^t\M_0^{1/2} ~ \mbox{for} ~ t \in [0,1].
\eeq
Note that $\M_t \in \PDH(\pdim)$ for $0 \le t \le 1$ and consequently $\PDH(\pdim)$ is said to form
a {\paino geodesically convex set}. A function $h: \PDH(\pdim) \rightarrow \R$ is then a {\paino geodesically convex function} if
\beq \label{eq:gconfun}
h(\M_t) \le (1-t)~h(\M_0) + t~h(\M_1) ~ \mbox{for} ~ t \in (0,1). 
\eeq
If the inequality is strict, then $h$ is said to be geodesically strictly convex. In the $\pdim =1$ dimensional real setting, geodesic convexity/strict
convexity is equivalent to the function $h(e^x)$ being convex/strictly convex in $x \in \R$. Thus, Condition~\ref{cond:rho} presumes $\rho(t)$ to be
geodesically convex. 

The concept of geodesic convexity enjoys properties similar to those of convexity in complex Euclidean space.  In particular, if $h$ is geodesically convex 
on $\PDH(\pdim)$ than any local minimum is a global minimum. Furthermore, if a minimum is obtained in $\PDH(\pdim)$ then the set of all minimums form a 
geodesically convex subset of $\PDH(\pdim)$.  If $h$ is geodesically strictly convex and a minimum is obtained in $\PDH(\pdim)$, then it is a unique minimum.

The following key result is given in \cite{zhang_etal:2013} for real positive definite symmetric matrices, although it also holds for
$\PDH(\pdim)$. We omit the proof for the complex case since it is analogous to the proof for the real case given in \cite{zhang_etal:2013}.

\begin{lemma} \label{lem:Mconvex}  
If  $\rho(t)$ satisfies  Condition~\ref{cond:rho}, then  
the cost function $\mathcal L(\M)$ in \eqref{eq:cost} is geodesically convex  in $\M \in \PDH(\pdim)$. In addition, if $r(x)$ is
strictly convex and $span\{\z_1, \ldots, \z_n\} = \C^{\pdim}$, then $\mathcal L(\M)$ is geodesically strictly convex in $\M \in \PDH(\pdim)$.
\end{lemma}

Recall that when using the notion of convexity in complex Euclidean space the cost function  $\mathcal L(\M)$
is convex in $\M^{-1} \in \PDH(\pdim)$, but not in $\M \in \PDH(\pdim)$, whenever $\rho(t)$ is a convex function. This includes the
well studied Gaussian case $\rho(t) =t$.   As shown below, geodesic convexity has the interesting property that if $\mathcal L(\M)$ being geodesically convex 
in $\M \in \PDH(\pdim)$ the it is also geodesically convex in $\M^{-1} \in \PDH(\pdim)$. 

From lemma \ref{lem:Mconvex}, we readily obtain the following corollary, which follows since the sum of two geodesically convex functions is easily seen to
be geodesically convex, and the sum of a geodesically  convex function and a geodesically strictly convex function is geodesically strictly convex.

\begin{corollary} \label{corr:penMconvex}
For $\rho(t)$ satisfying Condition \ref{cond:rho}, if $\mathcal P(\M)$ is geodesically convex/strictly convex in
$\M \in \PDH(\pdim)$, then the penalized cost function $\mathcal L_{\al,\be}(\M)$ in \eqref{eq:gpenfun} is geodesically convex/strictly convex
in $\M \in \PDH(\pdim)$ respectively.
\end{corollary}

As Lemma~\ref{lem:pencvx} below shows, Corollary~\ref{corr:penMconvex} applies to the penalty function of interest here, i.e., \ to $\mathcal P^*(\M) = \tr( \M^{-1})$.
Before proceeding, some further results and notations are reviewed. For Hermitian matrices $\A$ and $\B$ of the same order, the partial
ordering $\A \le \B$ or $\A < \B$ holds if and only if $\B-\A$ is positive semi-definite or positive definite, respectively. 
 The matrix $\M_{1/2}$ can be viewed as the geometric mean of $\M_0$ and $\M_1$ \cite{sra_hosseini:2013}, and as in the case of 
positive real numbers, it is known to be less than the arithmetic mean in the following sense,
\beq \label{eq:geo_arith}
\M_{1/2} \le (\M_0 + \M_1)/2, 
\eeq
with equality holding if and only if $\M_0 = \M_1$. 
It readily follows from its definition \eqref{eq:geopath} that for
$\K = \M^{-1}$ 
\beq \label{eq:inv_inv}
\K_t = \K_0^{1/2}\left(\K_0^{-1/2}\K_1\K_0^{-1/2}\right)^t\K_0^{1/2} = \M_t^{-1},
\eeq
and consequently \eqref{eq:geo_arith} also holds to $\M^{-1}$.  Equation \eqref{eq:inv_inv} together with the definition
of geodesic convexity shows that geodesic convexity in $\M$ implies geodesic convexity in $\M^{-1}$. 

Taking the trace on both side of \eqref{eq:geo_arith} when applied to $\M^{-1}$ then gives 
\[ \tr(\M_{1/2}^{-1}) < \left\{\tr(\M_0^{-1}) + \tr(\M_1^{-1})\right\}/2, \]
for $\M_0 \ne \M_1$.  That is, $\tr(\M^{-1})$ is midpoint geodesically strictly convex in $\M$. As with convex functions, midpoint geodesic strict convexity 
along with $\tr(\M^{-1})$ being continuous in $\M \in \PDH(\pdim)$ is sufficient to imply geodesically strict convexity and hence we obtain our desired result.

\begin{lemma} \label{lem:pencvx} The penalty term $\mathcal P^*(\M)=\tr(\M^{-1})$ is geodesically strictly convex in $\M \in \PDH(\pdim)$. 
\end{lemma}

Another interesting geodesically convex penalty function was  proposed by Wiesel \cite[Proposition 3]{wiesel:2012b}. Wiesel's  penalty has a specific property of being  scale invariant.  
To this point, it has been shown that under the stated conditions on $\rho$, the regularized loss function \eqref{eq:penfun} is geodesically
strictly convex.  To show that it has a unique minimum in $\PDH(\pdim)$, and consequently to show the regularized $M$-estimating equation \eqref{eq:penMest} 
admits a unique solution, it only needs to be shown that the minimum of \eqref{eq:penfun} occurs in the interior of $\PDH(\pdim)$. The following lemma 
shows that this holds and consequently implies the subsequent theorem.

\begin{lemma} \label{lem:boundary}
If $\rho(t)$ is bounded below, then $L^*_{\al,\be} (\M) \rightarrow \infty$ as $\M \rightarrow \partial \PDH(\pdim)$, i.e. the boundary of $\PDH(\pdim)$.
\end{lemma}

\proof Since $\rho(t)$ is bounded below, it only needs to be shown that if $\M \rightarrow \partial \PDH(\pdim)$ then
\[ -  \ln |\M^{-1}| + \al \tr(\M^{-1})  = \sum_{j=1}^\pdim \left( \frac{\al}{\lam_j(\M)} +\ln \lam_j(\M)\right) \rightarrow \infty. \]
However, $\M \rightarrow \partial \PDH(\pdim)$ if and only if $\lam_1(\M) \rightarrow \infty$ and/or $\lam_\pdim(\M) \rightarrow 0$. In either
case, $\al/\lam + \ln \lam \rightarrow \infty$ and so the lemma is established. \endproof

\begin{theorem} \label{thm:unique}
If $\rho(t)$ is bounded below and satisfies Condition~\ref{cond:rho}, then the penalized cost function \eqref{eq:penfun} 
has a unique minimum in $\PDH(\pdim)$. Furthermore, if $\rho(t)$ is also differentiable, then the minimum corresponds to the unique solution 
$\hat \M \in \PDH(\pdim)$ to the regularized $M$-estimating equation \eqref{eq:penMest}.
\end{theorem}

It is important to note that the existence and uniqueness of the regularized $M$-estimates do not require any conditions to be placed on 
the sample $\z_1, \ldots, \z_n$ for any $n \ge 1$. In particular, they exist and are unique for sparse samples, i.e.\  when $p < n$. This
is in constrast to the non-regularized $M$-estimates which requires a bound on the proportion of the data that can lie in any subspace
\cite{kent_tyler:1996}.

\section{Regularized Tyler's M-estimator} \label{sec:tylM}

An important case for which Lemma~Ê\ref{lem:boundary} and Theorem~\ref{thm:unique} do not hold is the regularized Tyler's $M$-estimator since in
this case $\rho(t) = \pdim \ln t$ is not bounded below. Hence this case requires special treatment. 

\begin{theorem} \label{thm:tyler}
For $\rho(t) = \pdim \ln t,  \alpha > 0$ and $ 0 \le \be < 1/\pdim$, the penalized cost function  $\mathcal L^*_{\al,\be} (\M)$ in \eqref{eq:penfun}   has a unique minimum 
in $\PDH(\pdim)$, with the minimum being obtained at the unique solution $\hat \M \in \PDH(\pdim)$ to \eqref{eq:Tyl}.
\end{theorem}

\proof 
Since $\z_i^\hop \M^{-1} \z_i  \ge \z_i^\hop \z_i / \lam_1(\M)$, it follows that 
\[
L^*_{\al,\be} (\M) \ge  C - \pdim \be \ln \lam_1(\M) + \sum_{j=1}^\pdim \left( \frac{\al}{\lam_j(\M)} +\ln \lam_j(\M)\right),
\]
where $C =  \frac{\pdim \be}{n}  \sum_{i=1}^\ndim \ln( \z^\hop_i \z_i )$ does not depend on $\M$. Again, the lemma follows since for
any $c > 0$, $\al/\lam + c \ln \lam \rightarrow \infty$ as $\lam \rightarrow 0$ or as $\lam \rightarrow \infty$. 
\endproof

Theorem~\ref{thm:tyler}  does not require any condition on the sample. However, to extend this result to
$1/p \le \be \le 1$, the following Condition A is sufficient and the following Condition B is necessary. These conditions 
holds for $n/\pdim > \be$ whenever the sample is in ``general position'', which occurs with probability one when
sampling from a continuous complex multivariate distribution. Note that the sufficient Condition A and the
necessary Condition B only differ when equality in the conditions is possible. %(This then implies Frederic's result.)

\vspace*{.2cm}
\noindent
{\bf Condition A.} For any subspace $\mV$ of $\C^{\pdim}$, $1 \le dim(\mV) < \pdim$, the inequality 
$\frac{ \# \{ \z_i \in \mV \}}{n} < \frac{dim(\mV)}{\pdim \be}$ holds. 
\vspace*{.2cm}

\noindent
{\bf Condition B.} For any subspace $\mV$ of $\C^{\pdim}$, $1 \le dim(\mV) < \pdim$, the inequality 
$\frac{ \# \{ \z_i \in \mV \}}{n} \le \frac{dim(\mV)}{\pdim \be}$ holds. 
\vspace*{.2cm}

%\noindent
%\textbf{Lemma 5G.}

We then have the following general result, the proof of which can be found in the Appendix. 

\begin{theorem} \label{thm:tyler2}
Suppose $\rho(t) = \pdim \ln t, \alpha > 0$ and $ 0 \le \be < 1$. 
\begin{itemize}
\item[a)] If condition A holds, then  \eqref{eq:penfun}  has a unique minimum 
in $\PDH(\pdim)$, with the minimum being obtained at the unique solution $\hat \M \in \PDH(\pdim)$ to \eqref{eq:Tyl}. 
\item[b)] If condition B does not hold, then \eqref{eq:penfun} does not have a minimum 
in $\PDH(\pdim)$, and \eqref{eq:Tyl} has no solution in $\PDH(\pdim)$.
\end{itemize}
\end{theorem} 

Note that if $\hat \M^*$ is a solution to \eqref{eq:Tyl} when using the shrinkage parameters $(\al,1-\al)$, i.e., the regularized Tyler's $M$-estimator with
$\tr(\hat \M^{-1})=\pdim$, then the solution to \eqref{eq:Tyl} when using $(\al,\be)$ is just a scaled multiple of $\hat \M^*$, namely 
\beq \label{eq:Tyl_eq}
\hat \M=[\be/(1-\al)] \cdot \hat \M^*.
\eeq 
So, when the main interest is on estimation of the covariance matrix or scatter matrix parameter up to a scale, as is the case in most applications, one can consider 
without loss of generality  (w.l.o.g.) the regularized Tyler's $M$-estimator with $\be=1-\al$. This existence and uniqueness of the regularized Tyler's $M$-estimator 
for this case, i.e., when $\be=1-\al$, has also been established in \cite{pascal_etal:2013}, but only under the condition that the data are in general position and hence
Conditions A and B are automatically satisfied for such samples.

A related regularized $M$-Tyler's estimator is given in \cite{chen_etal:2011} as the limit of the algorithm
\begin{align*}
\M_{k+1} &\leftarrow (1-\al) \frac{\pdim}{\ndim} \sum_{i=1}^\ndim \frac{\z_i \z_i^\hop}{\z_i^\hop \V_{k}^{-1} \z_i} + \al \I \\ 
\V_{k+1} &\leftarrow  \pdim \M_{k+1}/\tr(\M_{k+1}),
\end{align*}
where $\al \in (0,1)$ is a fixed regularization parameter. This algorithm represents a diagonally loaded (DL) version of the fixed-point algorithm given for Tyler's $M$-estimator. 
It was shown in \cite{chen_etal:2011} that the recursive algorithm above converges to a unique solution, referred to as CWH estimator, regardless of the initialization. Here, convergence means convergence in $\V_k$ and not necessarily in $\M_k$. It is not clear whether this estimator can be derived as a solution to a penalized cost function.

%Regularized  Tyler's $M$ estimator can be  computed using the iterative algorithm \eqref{eq:algor} (given it converges) with $\sigw(t)=p/t$. 

\subsection{Estimation of the regularization parameter} 

Let us define a scale measure of $\M \in \PDH_\pdim$ as 
\beq \label{eq:scale} 
%\tau(\M)=\tr(\M)/\pdim
\tau(\M)=\pdim/\tr(\M^{-1})
\eeq 
and $\V = \M/\tau(\M)$ as the respective shape matrix (thus verifying  $\tr(\V^{-1})=\pdim$).  Note that the regularized Tyler's $M$-estimator $\hat \M$ using $\be=1-\al$ 
can be considered as an estimator of shape matrix $\V$ as it verifies $\tr(\hat \M^{-1})=\pdim$).  
We now focus on this particular estimator and derive an oracle estimator of the shrinkage parameter $\al$ using a MSE criterion for similarity in shape.  We wish to emphasize that due to property 
\eqref{eq:Tyl_eq}, a regularized Tyler's $M$-estimator for general choice of $\be$ value (but fixed $\al$) is estimating the same shape matrix as the obtained solutions  will be proportional to each other. Thus in problems where an estimate of the scatter matrix (or covariance matrix) is only required up to a scale, one can rather see it as a problem for estimating the shape matrix. 

% (as only the scale is changed). 

%The oracle estimator  $\al_0$ of $\al$  assumes known shape matrix $\V$ and finds $\al_0$ as the minimizer of  
%$\E \big[ \| \V- \tilde \M_\al \|^2 \big]$.   

Since $\hat \M$ estimator in question is an estimator of shape matrix $\V$, one could aim at selecting $\al$ such that $\hat \M$  (or rather its approximation 
\eqref{eq:Mal}  for known $\V$) is as close as possible to $\V$ in the mean squared sense, i.e., $\E[ \| \Mal - \V\|^2]$.  This approach was used when deriving the oracle estimator 
of shrinkage parameter $\al$ for CWH estimator \cite{chen_etal:2011}. 
Alternatively, if we let $\Mn$ denote any matrix proportional to the true scatter matrix parameter $\Sig$, then  we should aim at choosing $\al$ such that 
$\Mn^{-1} \M_\al$ is as close as possible to being a scaled copy of an identity matrix, 
where  $\M_\al$ is {\paino clairvoyant estimator}  of $\hat \M$  given $\Mn$, defined as 
\beq \label{eq:Mal} 
\Mal = (1-\al) \frac{\pdim}{\ndim} \sum_{i=1}^\ndim \frac{\z_i \z_i^\hop}{\z_i^\hop \Mn^{-1} \z_i}   + \al \I, 
\eeq 
where w.l.o.g. we assume hereafter that $n_* = n$.
We then seek an oracle estimator $\al_o$ as  the minimizer of  the following MSE criterion 
\[
\al_o = \arg\min_\al  \E\big[ \|  \Mn^{-1} \Mal -  {\textstyle \frac{1}{\pdim} } \tr( \Mn^{-1} \Mal ) \I \|^2 \big] 
\]

%Thus the idea is similar as in \cite{chen_etal:2011} 

\begin{theorem}\label{th:oracle} The oracle estimator $\al_0$ when $\Mn$ verifies $\tr(\Mn^{-1})=\pdim$ 
is given by 
\begin{align} \label{eq:oracle}
 \al_{o} = \frac{ \pdim \tr(\Mn)  -1 } {\pdim \tr(\Mn) -1 + \ndim(\pdim+1)\{\pdim^{-1}\tr(\Mn^{-2}) -1\} }.  
 \end{align}
In the real case, the oracle estimator is
\[
 \al_{o,R} = \frac{ \pdim - 2 + \pdim \tr(\Mn) } {\pdim - 2 + \pdim \tr(\Mn) + \ndim(\pdim+2)\{\pdim^{-1}\tr(\Mn^{-2}) -1 \}} 
\]
\end{theorem} 

Since $\Mn$ is unknown, we estimate $\al_{o}$ in \eqref{eq:oracle} by simple plug-in estimate 
\beq \label{eq:oracle_est}
\hat{\al}_o = \frac{\pdim \tr(\hat \M)  -1} {\pdim \tr(\hat \M) -1 + \ndim(\pdim+1)\{\pdim^{-1}\tr(\hat \M^{-2}) -1\} }, 
\eeq 
where $\hat \M$ is  Tyler's $M$-estimator normalized to verify $\tr(\hat \M^{-1})=\pdim$ in the case that $n  \geq \pdim$.
In the cases that $n < \pdim$, one can employ a regularized Tyler's estimator with $\be < n/\pdim$ and $\al= 1- \be$.

\section{Numerical examples} \label{sec:simul} 

\subsection{Simulations study} 

%We consider the simulation set-up as in \cite{pascal_etal:2013} 
In our first simulation set-up, the covariance matrix is $\M$ %  = \tau(\M) \V$ 
is a real-valued correlation matrix (i.e., components $z_i$ have unit variances, real and imaginary parts are uncorrelated) of Toeplitz form
% whose elements  are determined by correlation coefficient $\rho$ as follows
\[
[\M]_{ij}  = \rho^{|i-j|},  \quad \rho \in (0,1) . 
\]
%We denote the  corresponding shape matrix as $\V = \M/\tau(\M)$, where $\tau(\M)$ is the scale measure \eqref{eq:scale}. 
Note that when $\rho$ is close to $0$, then $\M$ is 
close to an identity matrix and when $\rho$  tends to $1$, $\M$ tends a singular matrix of rank 1. % $\bo 1 \bo 1^\top$  
%To assess the performance of the estimators, we use the normalzied mean squared errror (NMSE) which is equal to 
%$ \| \hat \M - \M \|^2/\| \M|^2$, when $\hat \M$ is the CWH estimator and $\| \hat \M - \V \|^2/\| \V\|^2$, when $\hat \M$ is  the regularized 
%Tyler's $M$-estimator using $\be=1-\al$.  Note that $\tr(\M)=\pdim$, so CWH estimator is estimator of $\M$ in this case.
To assess the performance of the estimators, we use the distance measure  
\[
\dist^2 \equiv \dist^2(\Mn,\hat \M) = \| \{\pdim/\tr(\Mn^{-1} \hat \M)\} \, \Mn^{-1} \hat \M - \I \|^2
\]
which measures the ability of the estimator $\hat \M$ to estimate the scatter matrix $\M$ up to its scale. Above $\Mn$ can be any matrix $\Mn$ proportional to $\M$ 
%(e.g., $\M$ itself or $\V$) as 
since the distance measure verifies $\dist^2(c_1 \M,c_2 \hat \M)=\dist^2( \M,\hat \M)$ for $c_1,c_2>0$ and $\dist^2 = 0$ if $\Mn \propto \hat \M$. 
Hence, without any loss of generality, we can set $\Mn=\M$.
In this simulation we consider the regularized Tyler's $M$-estimator with $\be = 1- \al$ and the CWH estimator. Note that $\be=1-\al$ can be selected due to the property 
\eqref{eq:Tyl_eq}. We also compare the results with the (non-regularized) Tyler's $M$-estimator.  The samples $\z_1,\ldots, \z_n$ are  generated  from 
$\C \mathcal N_\pdim(\bo 0,\M)$, where 
the dimension of the data is $\pdim = 12$ and the number $\ndim$ of samples is $\ndim = 24$ and $\ndim = 48$.  Note that the simulation results would be the same 
if we sampled from any centered CES distribution, including compound Gaussian distributions, since the distribution of $\z_i/\|\z_i\|$ is the same for any CES distribution.

%This estimator is uniquely defined only up to a scale, but used distance measure  is the same for 
%$\hat \M$ and for $c \hat \M$ as it measures only the ability of the estimator to estimate the "shape" of data cloud, but not its scale. 

\begin{figure}[!t]
\centerline{\includegraphics[width=0.37\textwidth]{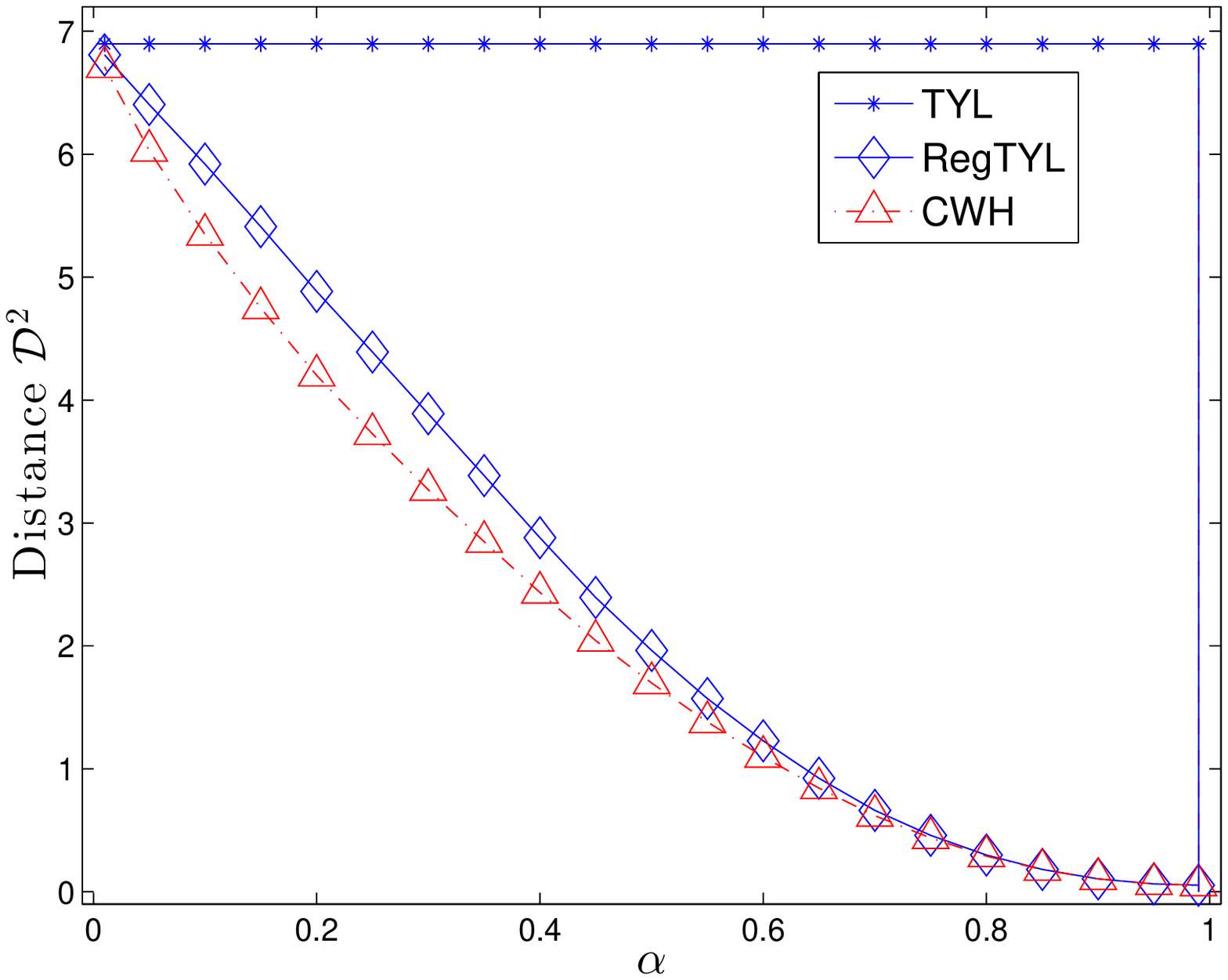}}
\centerline{(a) $\rho = 0.05$}
\centerline{\includegraphics[width=0.37\textwidth]{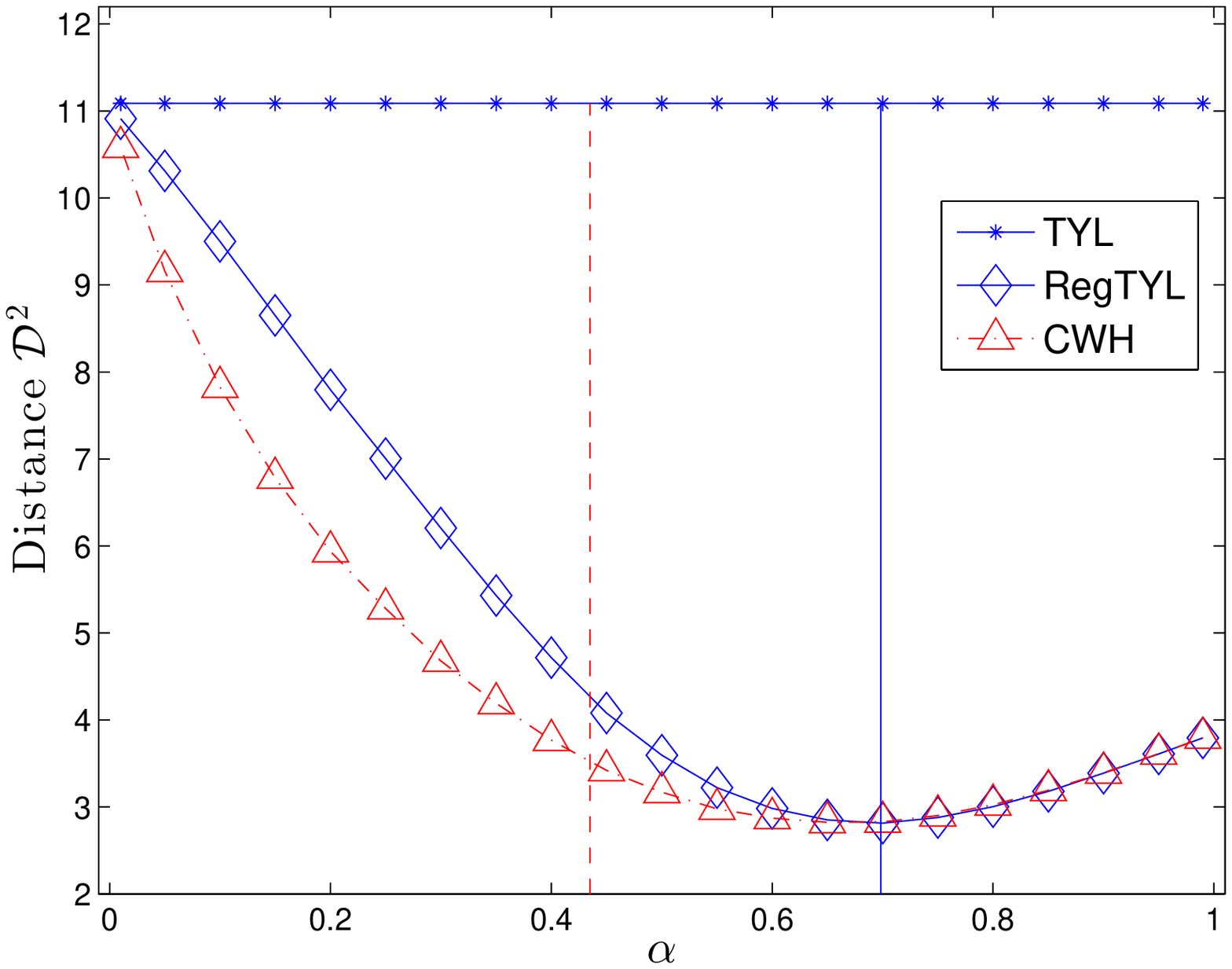}}
\centerline{(b) $\rho = 0.5$}
\centerline{\includegraphics[width=0.37\textwidth]{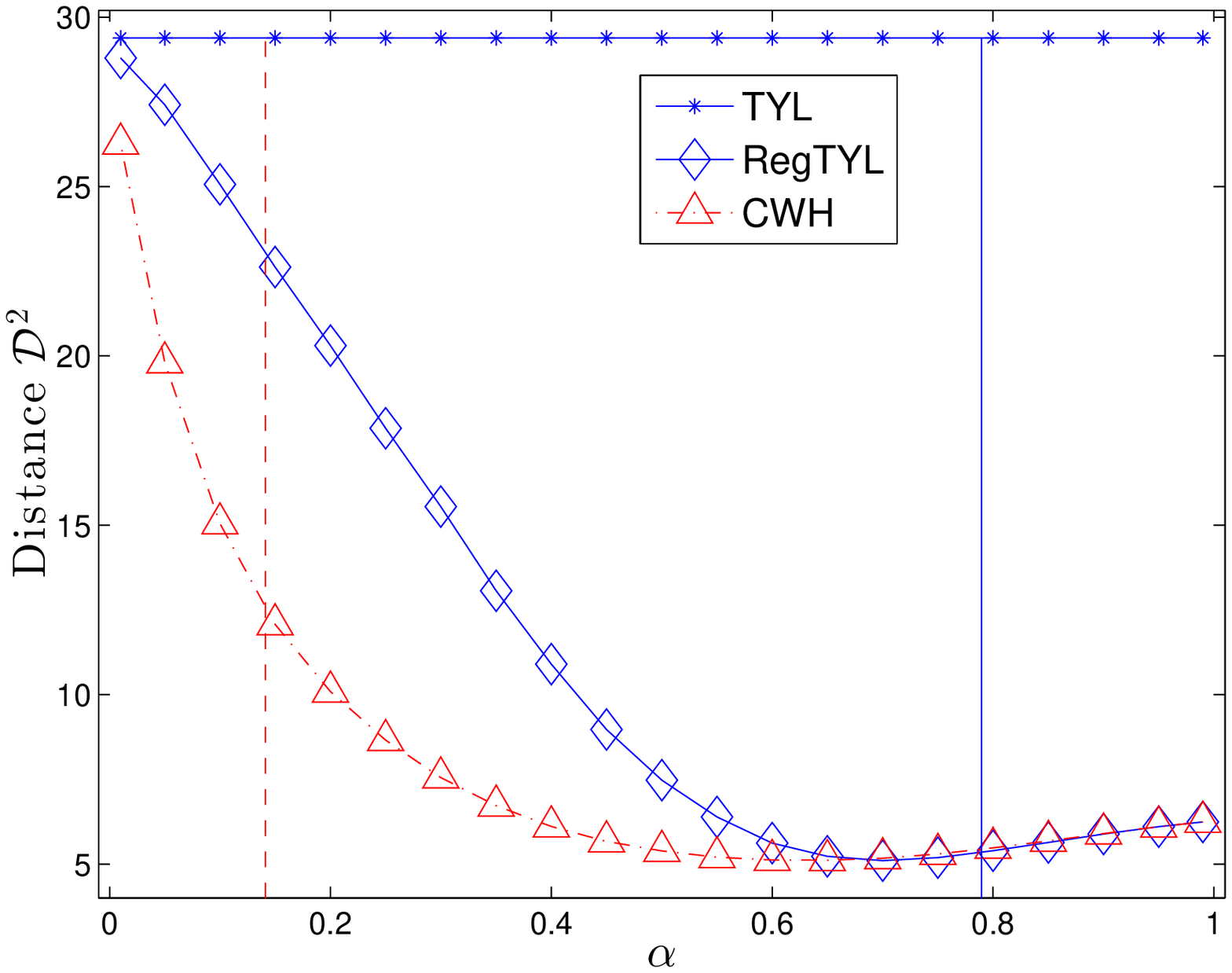}}
\centerline{(c) $\rho = 0.8$}
\vspace{-0.1cm}
\caption{ Distance $\dist^2$ of Tyler's $M$-estimator (TYL), regularized Tyler's $M$-estimator (RegTYL) and CWH estimator as a function of the shrinkage parameter $\al$. 
Results for different correlation matrix $\Sig$ given by $\rho=0.05, 0.5, 0.8$ are given from top to bottom. The dimension was $\pdim=12$, sample length was $\ndim=24$ 
and the results are averages of 1000 MC trials. The solid (resp. dotted) vertical line gives the oracle estimator $\al_0$ of RegTYL estimator in Theorem~\ref{th:oracle}   (resp. of CWH estimator in \cite[Theorem 3]{chen_etal:2011}).}
\label{fig:sim1re}
\end{figure}

 Figure~\ref{fig:sim1re} depicts the graphs of  $\dist^2$ averaged of $1000$ MC-trials as a function of shrinkage parameter $\al$ for CWH estimator, regularized Tyler's $M$-estimator (referred to as RegTYL)  and Tyler's $M$-estimator of scatter (referred to as TYL in the figure caption) in the cases that $\rho=0.01, 0.5, 0.8$ and the sample size is $n=24$. 
Figure~\ref{fig:sim1re} gives the results for sample length $n=48$. In both figures, the solid vertical line depicts the value of the oracle estimator $\al_{o}$ 
for  the regularized Tyler's $M$-estimator given by Theorem~\ref{th:oracle} and the dotted vertical line depicts the value of the oracle estimator 
$\al_o^{\mbox{\tiny CWH}}$ of  CWH estimator given  by \cite[Theorem 3]{chen_etal:2011}.

\begin{figure}
\centerline{\includegraphics[width=0.37\textwidth]{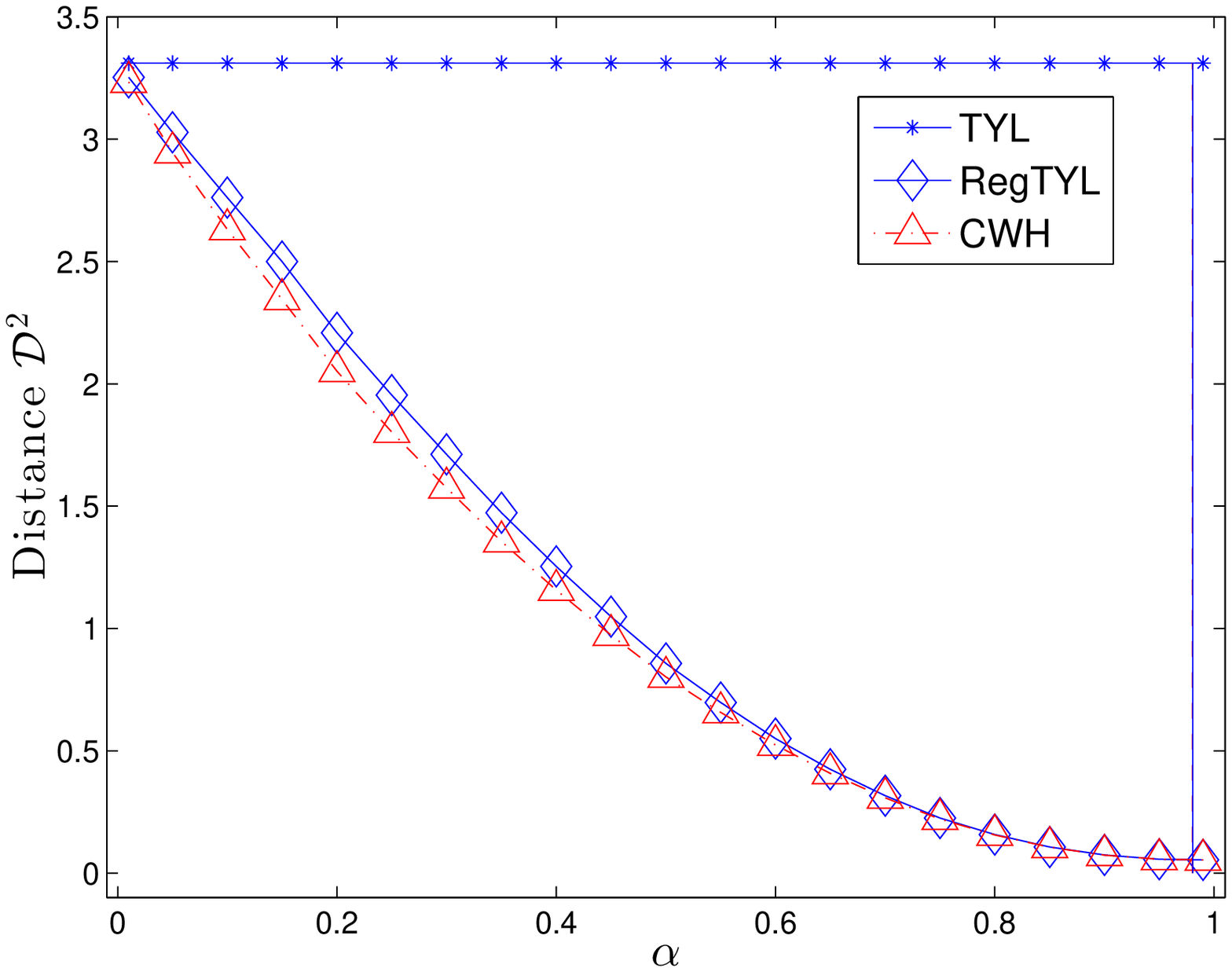}}
\centerline{(a) $\rho = 0.05$}
\centerline{\includegraphics[width=0.37\textwidth]{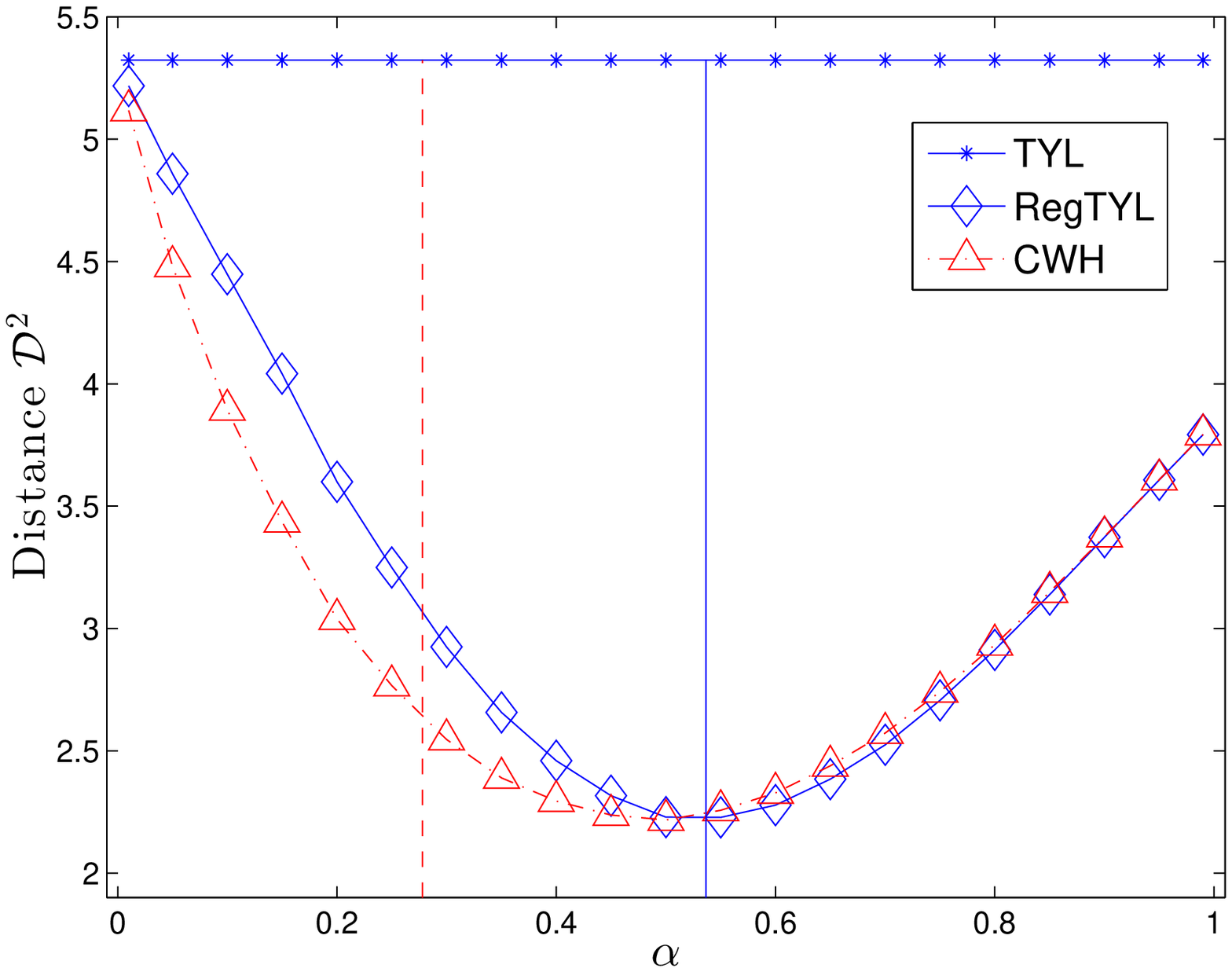}}
\centerline{(b) $\rho = 0.5$}
\centerline{\includegraphics[width=0.37\textwidth]{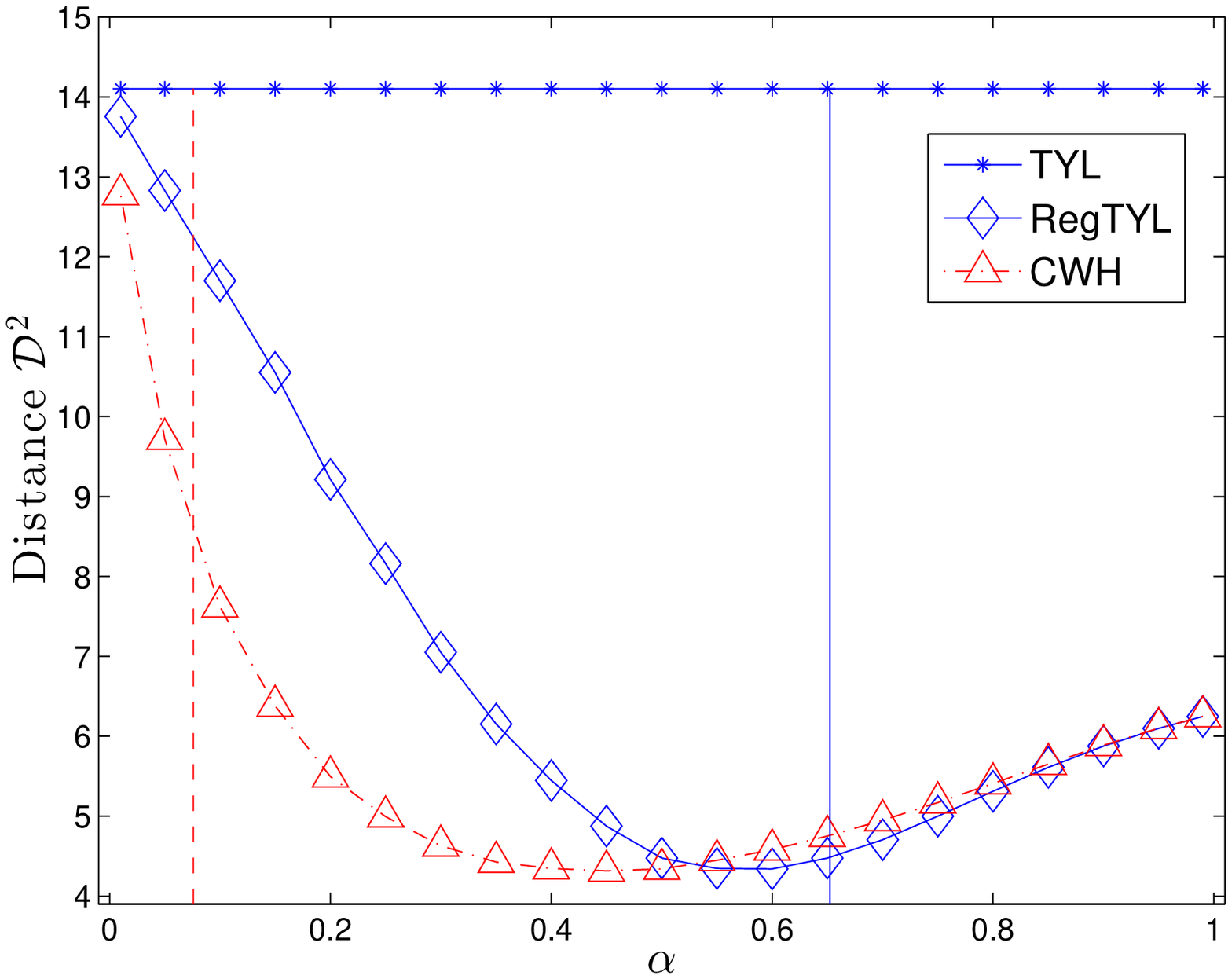}}
\centerline{(c) $\rho = 0.8$}
\vspace{-0.1cm}
\caption{ Distance $\dist^2$ %of Tyler's $M$-estimator (TYL), regularized Tyler's $M$-estimator (RegTYL) and CWH estimator 
for shrinkage estimators RegTYL and CWH 
as a function of the shrinkage parameter $\al$. 
Set-up is as in Figure~\ref{fig:sim1re}, but the sample size is twice larger $n=48$. }
%The solid (resp. dotted) vertical line gives the oracle estimator $\al_0$ of RegTYL estimator in Theorem~\ref{th:oracle}   (resp. of CWH estimator in \cite[Theorem 3]{chen_etal:2011}).}
\label{fig:sim1re2}
\end{figure}

%Three important remarks that these simulation   results indicate. 
The simulation   results indicate the following. 
First, the regularized Tyler's $M$-estimator (RegTYL) can be viewed as a generalization of 
Tyler's $M$-estimator since as $\al \to 0$ its performance tends to the performance of Tyler's $M$-estimator. This fact was also illustrated in % also in the simulations of 
\cite{pascal_etal:2013}.  For $\al \approx 0$,  the performance of the CWH estimator can still be quite different from that of Tyler's $M$-estimator. 
Second, the shape distance curves are very different for RegTYL and CWH estimators for the cases $\rho=0.5$ and $\rho=0.8$. 
Only for the case $\rho=0.05$ (i.e., when $\M$ is close to identity matrix) are they similar. In general, though, the value of $\al$ play a different role in RegTYL 
and CWH, and so comparing the two estimators for the same $\al$ is not particularly meaningful. 
Third, of primary interest is the performance of the oracle estimators for RegTYL, obtained at $\al_o$, and the performance of the CWH oracle estimator, obtained at
say $\al_0^{\mbox{\tiny CWH}}$. The figures illustrate that these two shrinkage generalizations of Tyler's scatter matrix provide fairly different estimators of scatter matrix,
and that RegTYL oracle estimator outperforms the CWH oracle estimator (when $\dist^2$ is used as a criterion).  In all cases, the shrinkage estimators (RegTYL and CHW) outperform the (non-regularized) Tyler's 
$M$-estimator (TYL).  For the case $\rho=0.05$ (i.e., $\M$ is being close to an identity matrix), both of the oracle estimators are close to being one (i.e., $\al_{0} \approx 1$ and $\al_0^{\mbox{\tiny CWH}} \approx 1$) as expected, i.e., both  estimators are being shrinked towards a scaled identity matrix.

\subsection{Radar detection using normalized matched filter} 

We address the problem of detecting a known %(transmitted) 
complex {\paino signal vector} (target response) 
$\p$ in {\paino received data} $\z = \gam \p + \c$,   where 
$\c$ represents the unobserved complex  {\paino noise} (clutter) r.v. and 
$\gam \in \C$ is a signal parameter modeled as an unknown deterministic parameter 
or as a random variable depending on the application at hand. Both the signal vector, 
the noise and the received data are $\dim$-variate. 
In radar applications, for example, $\gam$ is a 
complex unknown parameter accounting for both channel propagation effect and target backscattering
and $\p$ is the transmitted known radar pulse vector. 
The signal-absent vs. signal-present problem can  then be expressed as 
\beq \label{eq:hypo}
H_0 : |\gam| = 0 \quad \mbox{ vs. } \quad H_1 : |\gam| > 0.
\eeq  
We assume that $\c$ follows a centered CES distribution with a positive definite hermitian (PDH) scatter matrix parameter 
$\M$.  % = \sigma \Sig$, where $\sigma = \tr(\M)/\dim>0$ represents the unknown {\paino noise level} (average clutter power) and 
%$\V$ the known {\paino shape matrix} (normalized covariance matrix). 
For this problem, we consider the {\paino normalized matched filter (NMF) detector} 
\beq \label{eq:detector}
\Lamm \equiv \Lamm(\z;\p,\M)=  \frac{|\p^H \M^{-1} \z|^2}{(\z^H \M^{-1}\z)(\p^H \M^{-1}\p)} 
\ \hypo \  \lambda   
\eeq 
which is also referred to as constant false alarm rate (CFAR) matched subspace detector (MSD)  \cite{scharf_friedlander:1994}, 
or LQ-GLRT \cite{gini_greco:2002}, etc.  
It is well known that the distribution of $\Lam$ under $H_0$ is
 $\mathrm{Beta}(1,\dim-1)$, i.e., it is distribution-free under the class of CES distributions  \cite{kraut_etal:2001,ollila_tyler:2012}. This fact is of great practical importance 
 because the detector is CFAR under various commonly used clutter models (including the $K$-distribution, $t$-distribution, inverse Gaussian distribution which all belong to the class 
 of CES distributions).  Thus, to obtain  a probability of false alarm (PFA) equal to a desired level $P_{\mathrm{FA}}$ (e.g., $P_{\mathrm{FA}}=0.01$), the rejection threshold $\lambda$ can be set as the $(1-P_{\mathrm{FA}})$th quantile of the $\mathrm{Beta}(1,\dim-1)$ distribution   
\begin{gather} \label{eq:PFAlam}  
P_{\mathrm{FA}}  = \Pr(\Lamm > \lambda \vert H_0) = (1- \lambda)^{\dim-1}   
\end{gather} 
or $\lambda = 1 - {P_{\mathrm{FA}}}^{1/(\dim-1)}$;  see e.g.\ \cite{ollila_tyler:2012}.

However, in practice $\M$ is unknown and an {\paino adaptive  NMF detector} $\hat \Lam$ is obtained by replacing $\M$ 
by its estimate $\hat \M$ as in  \cite{conte_etal:2002,gini_greco:2002,kraut_etal:2001,kraut_etal:2005}. 
Note that the detector requires $\M$ only up to a scale since $\Lam=\Lam(\z;\p,c\M)$ for all $c>0$ and thus an  
estimate of the scatter matrix $\M$ is required up to a scale. Tyler's $M$-estimator, often called as fixed point estimator (FPE) in radar community, has become a 
popular method to estimate the unknown scatter matrix $\M$.  In radar applications $\hat \M$ is computed from signal free (clutter only), 
but the sample size $n$ is rarely large compared to the dimension $\dim$ (LSS/ISS cases). 
The  adaptive NMF detector $\hat \Lam$ based on the sample covariance matrix or any $M$-estimator of scatter does not retain the CFAR property since 
an $M$-estimator $\hat \M$ (although consistent) can be a highly inaccurate estimator in LSS/ISS cases. Naturally, the probability of detection is severely affected as well. 
We now illustrate by simulations that the regularized Tyler's $M$-estimators with estimated $\hat \al_o$ is able to provide the same CFAR  property and 
probability of detection (PD) as the theoretical  NMF that is based on the true scatter matrix $\M$. 

In our first simulation setting, we investigate how well the adaptive detector $\hat \Lam$ based on estimated $\M$ is able to main the preset PFA in \eqref{eq:PFAlam}. 
For each MC trial, the simulated data consist of received data $\z$ (used as input to NMF detector) and the secondary data $\z_1,\ldots,\z_n$ (used as input to estimate $\hat \M$).  
The data sets  are generated as i.i.d. random samples  from 
$\dim=8$ variate $K$-distribution $\C K_{\dim,\nu}(\bo 0,\M)$ with $\nu=4.5$.  
For $10000$ trials we calculated the empirical $P_{\mathrm{FA}}$ (the proportion of incorrect rejections) for a fixed  threshold $\lambda$ 
when the true scatter matrix $\M$  was generated randomly for each trial data set as follows. 
 We generated a random complex orthogonal $\pdim \times \pdim$ matrix 
$\P$ and a diagonal matrix $\bo D=\diag(d_1,\ldots,d_\pdim)$, where $d_i$'s were generated independently from  $Unif(0,1)$ distribution. Then the scatter matrix 
$\M$ was generated using the SVD as $\M= \P \bo D \P^\hop$. It should  be noted that the detector is invariant to the scale of $\M$, so the scale of $Unif(0,b)$ distribution of 
eigenvalues $d_i$ can be chosen to be $(0,1)$ without any loss of generality.  In our simulation we compare the following estimators of $\M$: 
\begin{itemize} 
\item TYL, referring to Tyler's $M$-estimator $\hat \M$. 
\item GLC,  referring to $\S_{\al,\be}$ in \eqref{eq:GLC}, where the parameters $\al$ and $\be$ are estimated as proposed in \cite[cf. Eq.'s (32) and (33)]{du_etal:2010}. 
\item RegTYL, referring to regularized Tyler's $M$-estimator of scatter with parameters  $\be=1-\hat \al_o$ and  $\al=\hat \al_o$,  $\hat \al_o$ given by \eqref{eq:oracle_est}.  
\item CWH estimator using the plug-in oracle estimator $\hat{\al}_o^{\mbox{\tiny CWH}}$ as proposed  in \cite[cf. Eq.'s (13) and (14)]{chen_etal:2011}. % based on (trace normalized) sample 
%sample covariance matrix of the samples \cite[in Eq.'s (14)]{chen_etal:2010} 
%\item Regularized Huber's $M$-estimator using $\be$ and $\al$ as for RegTYL above.  
%$\al_0$Êby \cite{
\end{itemize}

\begin{figure}[t]
\centerline{\includegraphics[width=7.6cm]{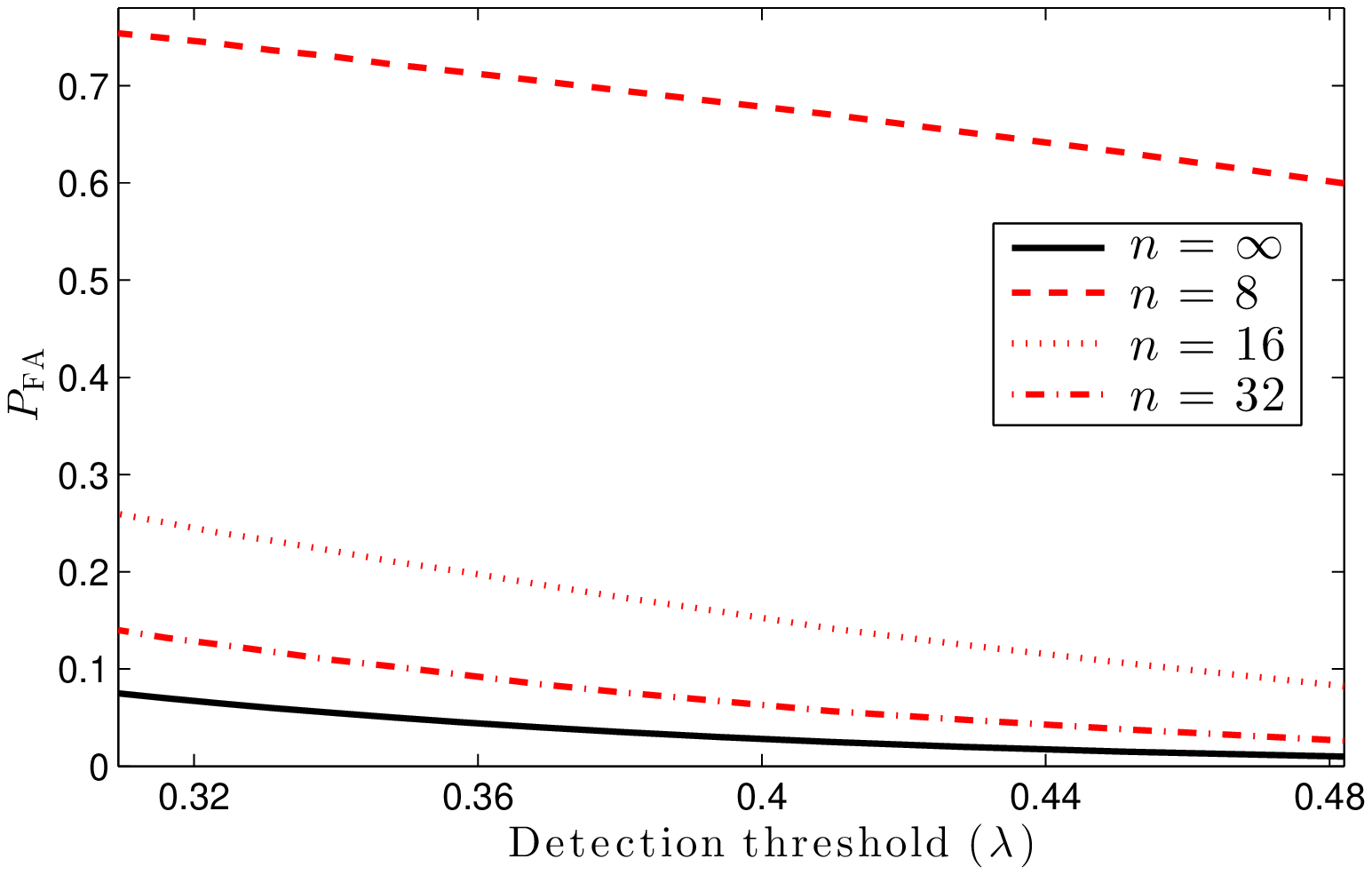}} 
{\scriptsize{\centerline{(a) TYL estimator } }}
\centerline{\includegraphics[width=7.6cm]{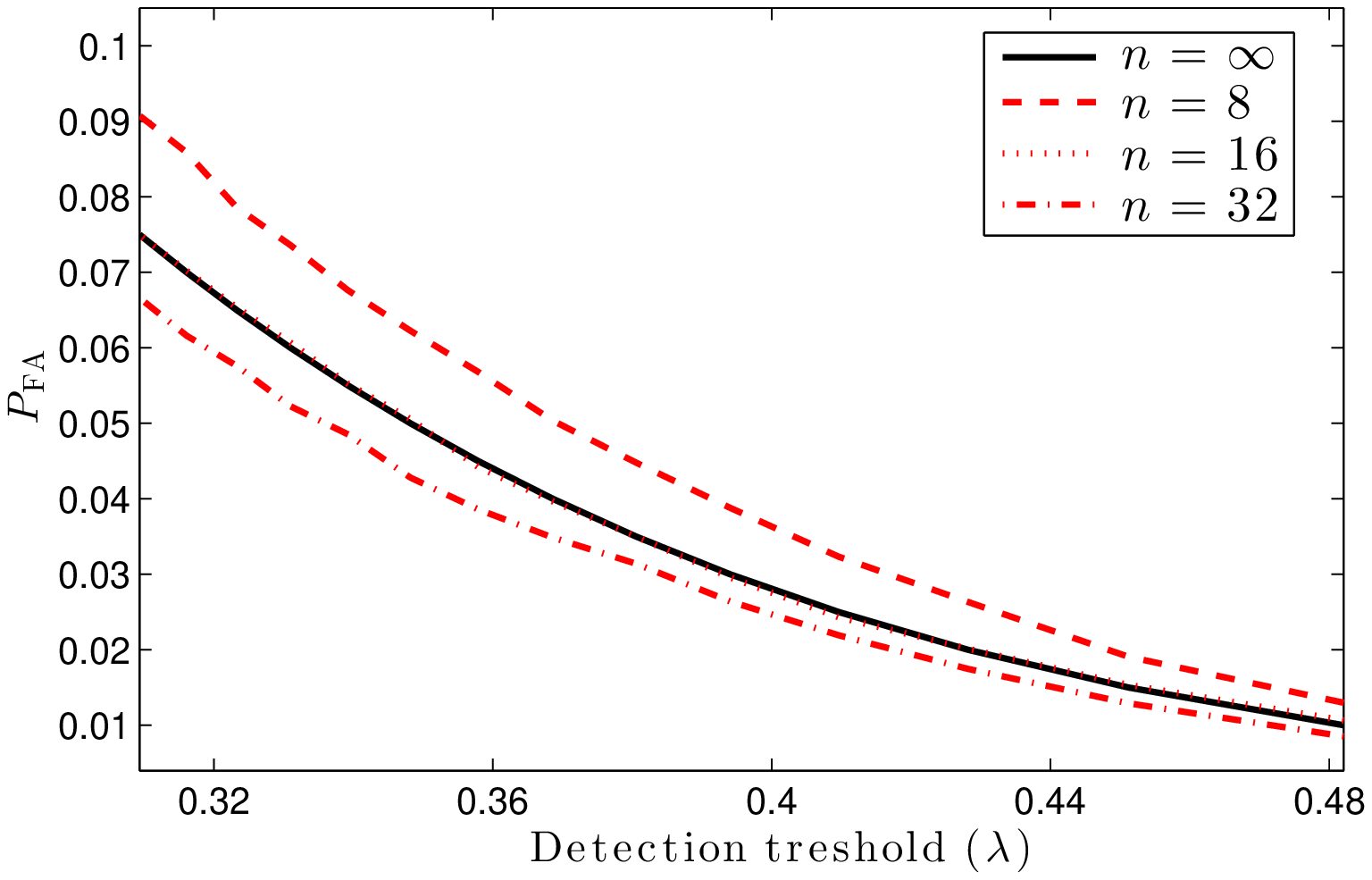}} 
{\scriptsize{\centerline{(b) GLC  estimator using $\hat \al$ and $\hat \be$} }}
\centerline{\includegraphics[width=7.6cm]{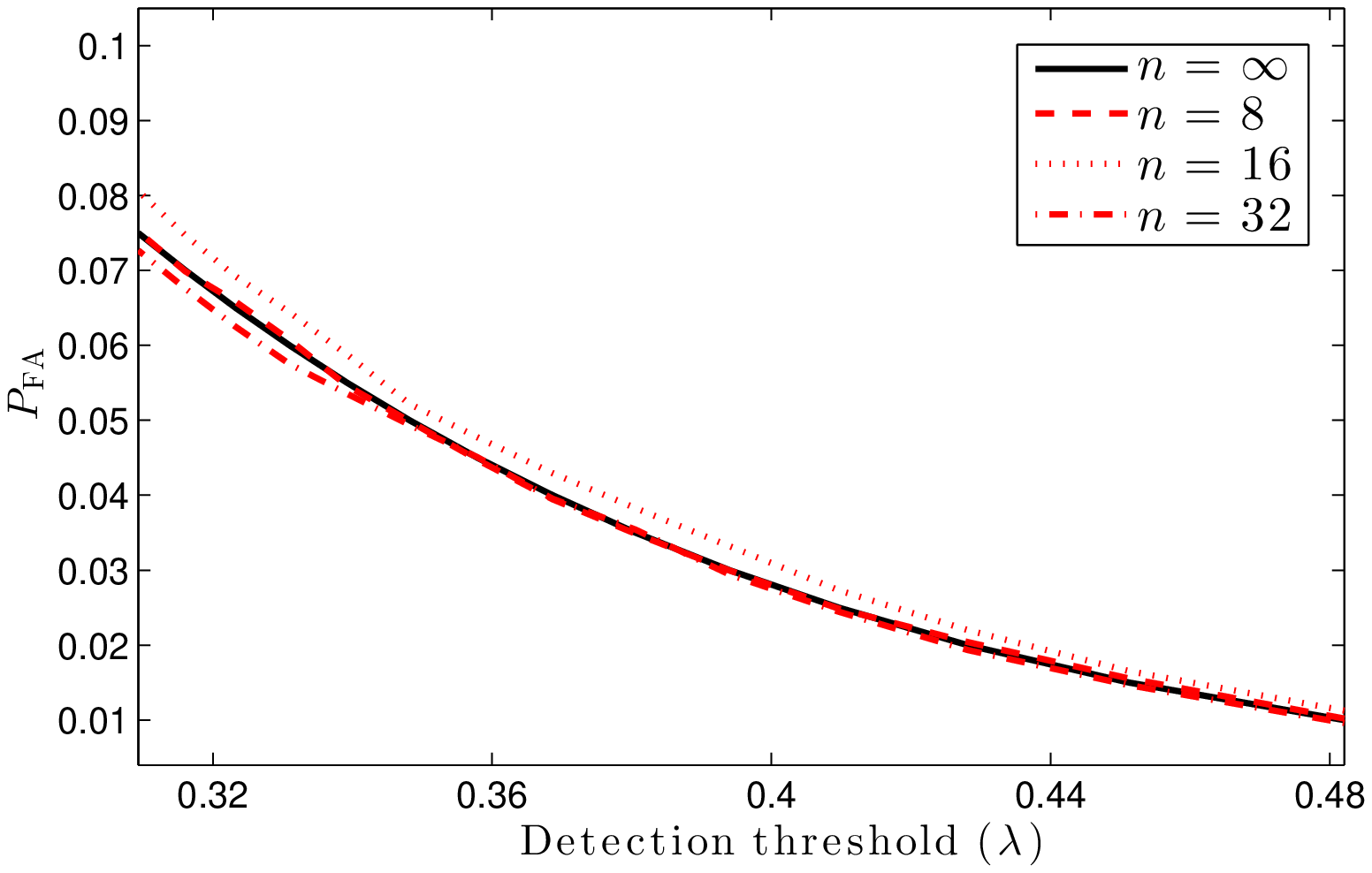}} 
{\scriptsize{\centerline{(c) Reg-TYL  estimator using $\hat \al_o$} }}
\centerline{\includegraphics[width=7.6cm]{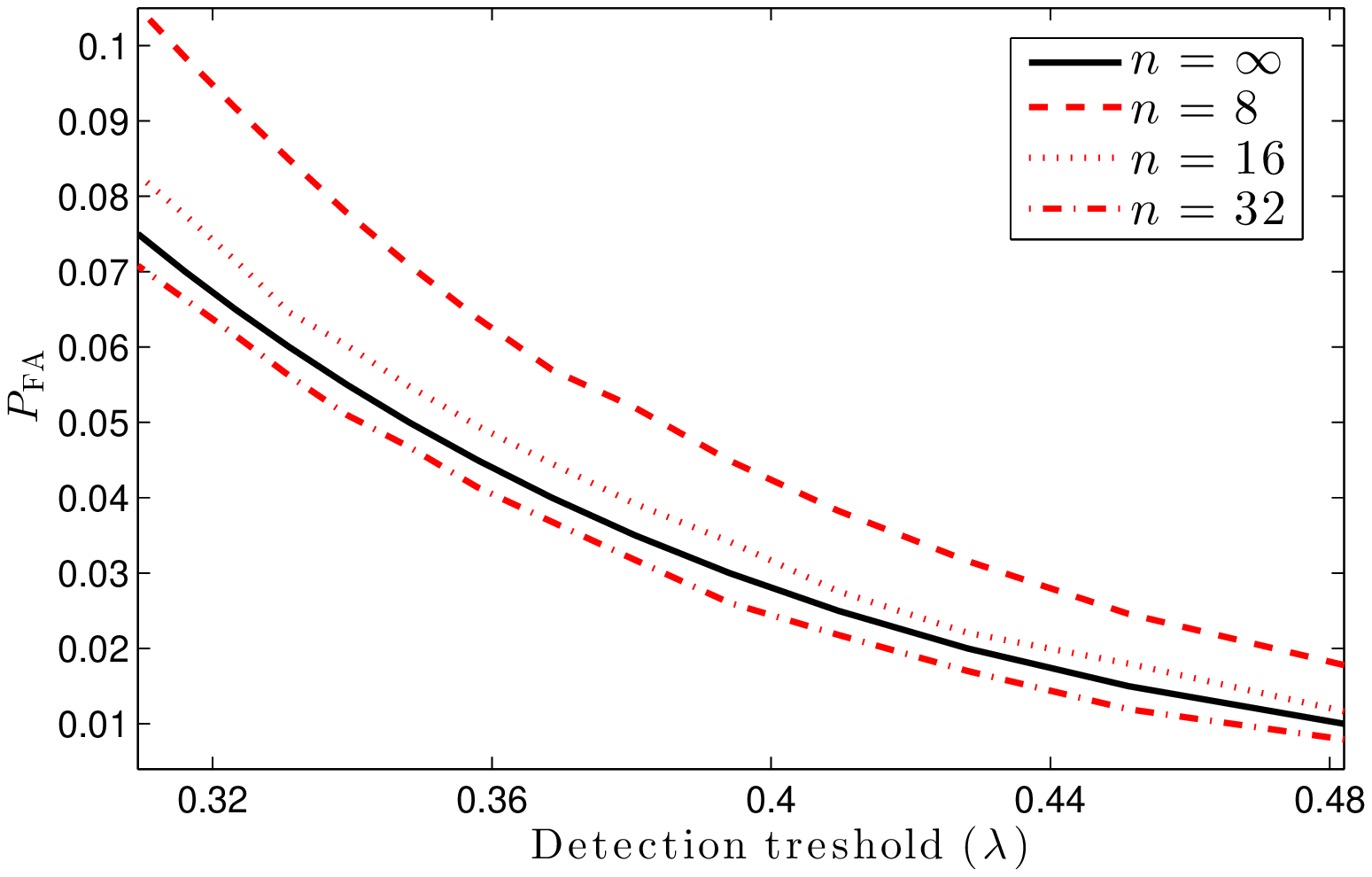}}
%\centerline{(d) CWH estimator using $\hat \al_o$} 
{\scriptsize{\centerline{(d) CWH estimator using $\hat \al_o^{\mbox{\tiny{CWH}}}$} }}
\caption{Empirical $P_{\mathrm{FA}}$ for adaptive detector employing different scatter matrix estimators 
under $K$-distributed clutter with $\nu = 4.5$ and different sample lengths $n$ of the secondary data.
The dimension $m=8$ and the clutter covariance matrix $\M$ was generated randomly for each 10000 trials.}  
\label{fig:PFAsim}
\end{figure}

Note that the shape parameter $\nu=4.5$ is large so that the $K$-distribution is close to being Gaussian. Namely, when $\nu$  descends towards zero, the $K$-distributions gets heavier tailed.  Since the $K$-distribution in question is not  heavy-tailed in nature, GLC estimator is expected to produce reliable estimates. 
This would not be the case for $\nu$ closer to 0. 
Figure~\ref{fig:PFAsim}  depicts empirical PFA curves of adaptive detectors.    
Note that the solid curve ($n=\infty$) corresponds to the theoretical PFA curve in \eqref{eq:PFAlam} for NMF $\Lamm$ with known $\M$.
As can be seen in Figure~\ref{fig:PFAsim}(a), when the detector is based on Tyler's $M$-estimator and the sample length is small $n=8,16,32$, there exists a remarkably huge gap between the observed  PFA and the desired (theoretical) PFA especially when the desired PFA is relative large (e.g., $P_{\mathrm{FA}}  = 0.05$).   
The performance of shrinkage estimators, GLC, RegTYL and CWH, depicted in Figures~\ref{fig:PFAsim}(b)--(d) illustrate their superior performance  compared to (non-regularized) Tyler's  $M$-estimator.   RegTYL estimator has clearly the best performance here: it is able to maintain the empirical PFA very close to the theoretical (desired) PFA for all sample lengths $\ndim=8,16,32$ considered. As can be seen, CWH estimator has second best performance but it is severely overestimating the true PFA when $\ndim=8$ and slightly underestimating for $\ndim=32$.   GLC estimator 
on other hand  has good performance only for the largest sample length $\ndim=32$ in which case there is a good match between the theoretical PFA and empirical PFA curves.  
%In radar applications it is common to set $P_{\mathrm{FA}} \leq 0.01$ in which case only a small 
%mismatch between the empirical PFA and the nominal PFA is observed.
Finally, it is important to recall again that the same graphs would be obtained (on the average) for the TYL, RegTYL and CWH estimators
if the simulation samples are drawn from any other CES distribution due to distribution-free property of these estimators. This is not true, though, for the GLC estimator whose 
performance depends on the underlying  CES distribution. Due to its inefficiency at longer tailed non-Gaussian distributions and vulnerability to outliers,  the GLC estimator 
can not be recommended in radar applications since the clutter is often heavy-tailed (spiky) in nature. If the shape parameter $\nu$Êof the $K$-distribution is 
close to zero, then the performance of GLC estimator degrades severely whereas the performance of RegTYL and CWH estimators remain unaffected.

\begin{figure}[!ht]
\centerline{\includegraphics[width=9.8cm]{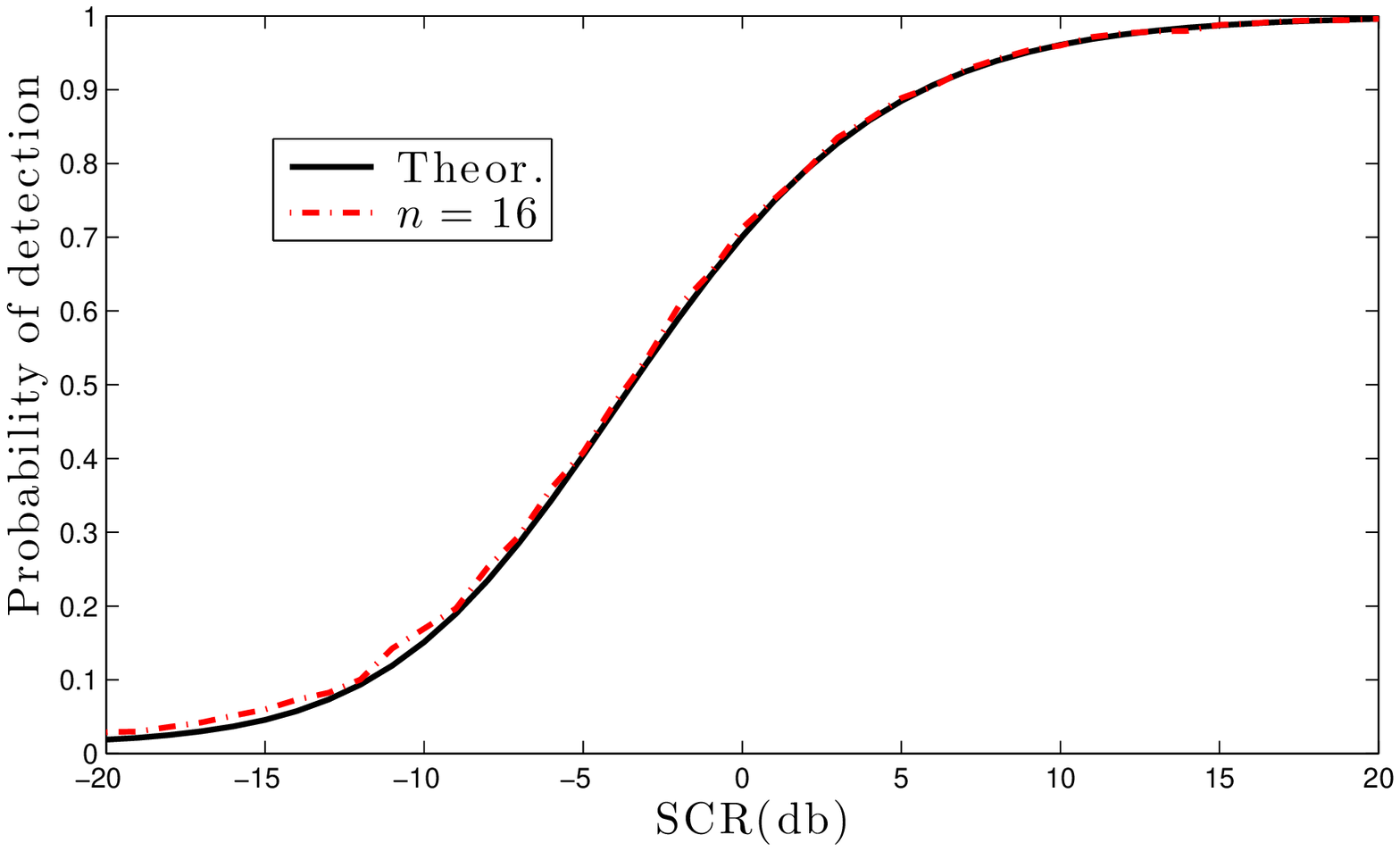}} 
%\vspace{-0.2cm}
\caption{Observed PD as a function of the SCR $\sigma^2_{|\gam|}/\sigma^2$ of the adaptive detector 
based on regularized Tyler's $M$-estimator computed from $\ndim=16$ secondary (signal free) data  
The clutter follows $\C K_{\pdim,\nu}(\bo 0,\M)$ distribution with $\nu = 4.5$ and $\M=\sigma^2 \I$. 
The dimension is $\pdim=8$, the pulse has  norm $\|\p\|^2 = \pdim$, signal amplitude $|\gam| \sim Rayl(\sigma_{|\gam|})$ 
and observed PD is  averaged of 5000 MC trials. The detection threshold $\lambda$  of the adaptive detector was set to give theoretical PFA 1\%.}
\label{fig:PDsim}
\end{figure}

In the second simulation study, we inspect the PD of the adaptive NMF detector. We only include the RegTYL estimator in this study since it had the best performance among the 
all considered estimators.    
Let us now assume (as in the Swerling-I target model) that under $H_1$ the signal amplitude   %$\alpha \sim \C N(0, \sigma^2_{\alpha})$ which implies that 
$|\gam |$ has a Rayleigh distribution with scale $\sigma_{|\gam|}$. %=\sigma_\al/\sqrt{2}$. We assume Rayleigh fluctuating amplitude $|\gam|$ and $K$-distributed clutter with the same shape parameter $\nu=4.5$. 
Then for each  MC-trial, the  data set consists of received data $\z$ generated from $H_1$ (and used as input to adaptive detector)  
and a sample of secondary data $\z_1,\ldots,\z_n$ from $H_0$ (and used to estimate $\M$ required by the adaptive detector). 
The scatter matrix parameter of the clutter is  $\Sig=\sigma^2 \I$, $\|\p\|^2 = \dim$ and 
the threshold $\lambda$ is set to give %as $0.01$th upper quantile of the $B(1,m-1)$ distribution which quarantees 
$P_{\mathrm{FA}}=0.01$. As can be noted from Figure~\ref{fig:PFAsim}(c), this threshold value also accurately reflects  the  observed (empirical) PFA of the adaptive detector. 
The theoretical PD curve of NMF statistics $\Lamm$ (based on true $\M$)
can be calculated numerically as a simple 1-dimensional 
integral \cite[Eq. (11)]{ollila_tyler:2012} for each fixed signal to clutter (SCR) ratio $\sigma^2_{|\gam|}/\sigma^2$ (dB).  
Figure~\ref{fig:PDsim} plots the theoretical PD curve as a function of the SCR %$\sigma^2_{|\gam|}/\sigma^2$ (dB)  
and the observed PD (the proportion of correct rejections) over 5000 simulated independent MC trials (for each fixed SCR $ = -20,-19,
\ldots,19,20$ (dB)). % for the adaptive detector $\hat \Lam$ utilizing Tyler's $M$-estimator. 
As can be seen the adaptive NMF detector based on RegTYL estimator is able to maintain accurately the true PD of the (theoretical) NMF detector. 
Results for sample length $n=16$ of the (signal-free) secondary data is $\ndim=16$ in our simulations.

\section{Conclusions} 

A general class of regularized $M$-estimators was proposed that constitute a natural generalization of $M$-estimators of scatter matrix by Maronna \cite{maronna:1976} 
% were proposed which 
%are suitable also 
%for low or insufficient sample support 
but are suitable also in small $\ndim$ and large $\pdim$ problems.  The considered  class was defined  as a solution to a penalized $M$-estimation cost function that depend on a pair $(\al,\be)$ of regularization parameters.  General conditions for uniqueness of the solution were established using the concept of geodesic convexity. For the regularized Tyler's $M$-estimator, necessary and sufficient conditions for uniqueness of the penalized Tyler's cost function were established separately  %and these were  and under more strict conditions on the sample, 
% These conditions were more general than in \cite{pascal_etal
%For the regularized Tyler's $M$-estimator, we also derive a simple, 
and a closed form (data dependent) choice for the 
regularization parameter was derived using the mean-squared error between shape matrices. 
An iterative algorithm  that was shown to converge  to the solution of the regularized $M$-estimating equation under general conditions was provided.  
Simulations studies and a radar detection example   illustrated the  usefullness of the proposed  methods.

\appendix

\section*{Proof of Theorem \ref{thm:tyler2}}

\proof a) Express $\Gamma = \M^{-1} = \gamma \bo M$ with $\tr(\bo M) = 1$, and so
$L^*_{\al,\be}(\M) = L_1(\gamma) + L_2(\bo M)$, where
\begin{align*}
L_1(\gamma) &= \pdim (\be - 1) \ln(\gamma)  + \al \gamma  \\
L_2(\bo M) &= \frac{\pdim \be}{n} \left\{\sum_{i=1}^n \ln (\z_i^\hop \bo M \z_i)\right\} - \ln | \bo M |.
\end{align*} 
Now if $\M \rightarrow \partial \PDH(\pdim)$ then either $\gamma \rightarrow 0$, $\gamma \rightarrow \infty$, or $\bo M \rightarrow \partial \PDH(\pdim)$.  If $\gamma$ goes to zero or infinity, it readily follows that $L_1(\gamma) \rightarrow \infty$ since for any $c > 0$,  $\al\gamma - c \ln \gamma \rightarrow \infty$ as $\gamma \rightarrow 0$ or 
as $\gamma \rightarrow \infty$. 

So, we only need to consider what happens to $L_2(\bo M)$ as $\bo M \rightarrow \partial \PDH(\pdim)$. Since the set of positive semi-definite Hermitian matrices with trace one is compact, it is sufficient to consider a sequence $\bo M_k \rightarrow \mM$, where $\mM$ is a singular positive semi-definite Hermitian matrix with
trace one.  Hence $ 1 < \mbox{rank}(\mM) < \pdim$. Let $\lam_1(\bo M) \ge \cdots \ge \lam_p(\bo M)$ denote the eigenvalue of $\bo M$.
Since eigenvalues are continuous functions, $\lam_j(\bo M_k) \rightarrow \lam_j(\mM)$.
The spectral value decomposition gives $\bo M_k = \sum_{j=1}^\pdim \lam_j(\bo M_k) \tha_{k,j}\tha_{k,j}^\hop$, where $\bo M_k \tha_{k,j} = \lam_j(\bo M_k) \tha_{k,j}$ 
with $\tha_{k,j}^\hop\tha_{k,m} = \delta_{j,m}$. By compactness, it can be assumed without loss of generality that $\tha_{k,j} \rightarrow \tha_j$, $j = 1, \ldots, \pdim$, with 
$\tha_{j}^\hop\tha_{m} = \delta_{j,m}$. For $j = 1, \ldots, \pdim$, let $S_j$ denote the subspace of $\C^\pdim$ spanned by  
$\{\tha_j, \ldots, \tha_\pdim\}$, $S_{\pdim+1} = \{ \0 \}$ and $D_j = S_j \backslash S_{j+1} = \{ \z \in \C^\pdim ~|~ \z \in S_j, \z \notin S_{j+1} \}$. Also, let
$n_j = {}\#\{\z_i \in D_j\}$ and $N_j = {}\#\{\z_i \in S_j\}$. 

For $n_j \ge 1$ and $\z_i \in D_j$, $\z_i^\hop \bo M_k \z_i \ge \lam_j(\bo M_k) | \tha_{k,j}^\hop \z_i |^2 \ge \lam_j(\bo M_k) c_{k,j}$, where 
\begin{align*}
c_{k,j} = \min\{| \tha_{k,j}^\hop \z_i |^2 ;& \z_i \in D_j\} \\ 
 & \rightarrow c_{j} = \min\{| \tha_{j}^\hop \z_i |^2 ; \z_i \in D_j\}  > 0.
\end{align*} 
For $n_j = 0$, let $c_{k,j} = c_j = 1$. Hence, 
\begin{align*}
L_2(\bo M_k) \ge \frac{\pdim \be}{n} & \sum_{j = 1}^\pdim n_j \ln(c_{k,j})  \\  & + \sum_{j = 1}^\pdim \left(\frac{\pdim \be n_j}{n} -1 \right) \ln\{\lam_j(\bo M_k)\}. 
\end{align*}
The first term on the right converges to $\frac{\pdim\be}{n} \sum_{j = 1}^\pdim n_j \ln(c_{j}) > -\infty$ and for $ j \le r = \mbox{rank}(\mM),  0 < \lam_j(\mM) < 1$. 
So, to complete the proof of part (a), it only needs to be shown that
\[
 L_3(\bo M_k) = \sum_{j = r+1}^\pdim \left(\frac{\pdim \be n_j}{n} -1 \right) \ln\{\lam_j(\bo M_k)\} \rightarrow \infty. 
\]
Condition A implies $\frac{p\be N_j}{n} < \pdim - j +1$ for $j = 2, \ldots \pdim$. Also, since $n_j = N_j - N_{j+1}$ with $N_{p+1} = 0$, it follows that
$\left(\frac{p\be n_j}{n} -1 \right)  < a_j$, where $a_j = \left(\pdim - j -\frac{p\be N_{j+1}}{n}\right)$ for $j = 2, \ldots, \pdim$. Condition A 
also insures that $a_j \le 0$ and so $\left(\frac{p\be n_j}{n} -1 \right)$ is strictly negative. Finally, for $j = r+1, \ldots, \pdim$,
$\ln\{\lam_j(\bo M_k)\} \rightarrow -\infty$. Thus, each term in $L_3(\bo M_k)$ must goes to $\infty$. 

b) If condition B does not hold, then there exists a subspace $\mV_o$ such that $\frac{ n_o}{n} > \frac{d_o}{\pdim \be}$,
where  $n_o = {}\# \{ \z_i \in \mV_o \}$ and $d_o = \mbox{dim}(\mV_o)$, with $1 \le d_o < \pdim$.  
Construct the sequence $\Gamma_k = \M_k^{-1} \in \PDH(\pdim)$ as follows. 
Let $\Gamma_k$ having eigenvalues $1$ and $\gamma_{k,o}$ with multiplicities $\pdim-d_o$ and $d_o$ respectively, with $\gamma_{k,o} \rightarrow 0$.
Also, for every $k$, let the eigenspace associated with $\gamma_{k,o}$ be $\mV_o$. Part (b) then follows by showing $L^*_{\al,\be}(\M_k) \rightarrow -\infty$. 

To show this, note that  $L^*_{\al,\be}(\M_k) = L_{a,k} + L_{o,k}$, where 
\[
L_{o,k} = \left(\frac{\pdim \be n_o}{n} - d_o \right) \ln(\gamma_{k,o}) \quad \mbox{and} 
\]
\[
L_{a,k} =  \frac{\pdim \be}{n} \left\{\sum_{\z_i \in \mV_o} \ln(\z_i^\hop\z_i) +  \sum_{\z_i \notin \mV_o} \ln(\z_i^\hop \Gamma_k \z_i)\right\} - \alpha \tr(\Gamma_k).
\]
It readily follows that $L_{a,k} \rightarrow L_a < \infty$. Also, $L_{o,k} \rightarrow -\infty$ since $\log(\gamma_{o,k}) \rightarrow -\infty$
and $\frac{\pdim \be n_o}{n} > d_o$. 
\endproof

\section*{Proof of convergence of algorithm \eqref{eq:algor}}

\proof Suppose $\rho(t)$ is continuously differentiable, satisfies Condition 1, and  $\wei(t) = \rho'(t)$ is non-increasing. Also, assume 
the M-estimating equation (7) has a unique solution. Conditions for uniqueness are given in Theorems 1, 2 and 3. 

Let $\hS$ be the unique solution to (7), and define $\bV_k = \hS^{-\eh}\hS_k\hS^{-\eh}$.
Algorithm (9) can then be re-expressed as 
\[
 \bV_{k+1} = G(\bV_k) \equiv \frac{\beta}{n} \sum_{i=1}^n \wei(\y_i^\hop \bV_k^{-1} \y_i) \y_i \y_i^\hop + \alpha \hS^{-1}, 
\]
where $\y_i =  \hS^{-\eh} \x_i$ for $i = 1, \dots, n$.  From (7), it follows that $G(\I_\pdim) = \I_\pdim$.  Note that $\bV_k \in \PDH(\pdim)$, and so let
$\lambda_{1,k} \ge \cdots \ge \lambda_{\pdim,k} > 0$ denote the eigenvalues of  $\bV_k$. The objective is to show 
that $\bV_k \to \I_\pdim$ as $k \to \infty$.

\begin{lemma} \label{B} \mbox{ } \\[-16pt]
\begin{itemize}
\item[{\rm (i)}]  $\lambda_{1,k} > 1 \Rightarrow  \lambda_{1,k+1} < \lambda_{1,k}$. \\[-16pt]
\item[{\rm (ii)}]  $\lambda_{1,k} \le 1 \Rightarrow  \lambda_{1,k+1} \le 1$. \\[-16pt]
\item[{\rm (iii)}]  $\lambda_{\pdim,k} < 1 \Rightarrow  \lambda_{\pdim,k+1} > \lambda_{\pdim,k}$. \\[-16pt]
\item[{\rm (iv)}]  $\lambda_{\pdim,k} \ge 1 \Rightarrow  \lambda_{1,k+1} \ge 1$.
\end{itemize}
\end{lemma}

\proof 
(i) Since $\wei(t)$ in non-increasing, and $\psi(t) = t\wei(t)$ is non-decreasing, it follows that
$\wei(\y^\hop \bV_k^{-1} \y) \le \wei(\y^\hop \y/\lambda_{1,k}) =  \lambda_{1,k} \psi(\y^\hop \y/\lambda_{1,k})/ \y^\hop \y \le \lambda_{1,k} \wei(\y^\hop \y)$, 
and so
\begin{align*}
\bV_{k+1} \le \lambda_{1,k} ~ \frac{\beta}{n} &\sum_{i=1}^n \wei(\y_i^\hop\y_i) \y_i \y_i^\hop + \alpha \hS^{-1}  \\ 
&= \lambda_{1,k} G(\I_\pdim) + (1-\lambda_{1,k}) \alpha \hS^{-1}. 
\end{align*}
Thus, $\bV_{k+1} <  \lambda_{1,k} G(\I_\pdim) =  \lambda_{1,k} \I_\pdim$, and so part (i) follows. 

(ii) Since $\wei(t)$ is non-increasing, $\wei(\y^\hop \bV_k^{-1} \y) \le \wei(\y^\hop \y/\lambda_{1,k}) \le \wei(\y^\hop \y)$. Consequently, 
$ \bV_{k+1}  \le G(\I_\pdim) = \I_\pdim$, and so part (ii) follows.

The proofs for part (iii) and  (iv) are analogous.
\endproof

\begin{lemma} \label{C} \mbox{ } \\[-16pt]
\begin{itemize}
\item[(i)]  $\limsup \lambda_{1,k} \le 1$. \\[-16pt]
\item[(ii)]  $\liminf \lambda_{\pdim,k} \ge 1$.
\end{itemize}
\end{lemma}  

\proof
The proof is by contradiction. To show part (i), presume $\lambda_1 \equiv \limsup \lambda_{1,k} > 1$. By Lemma \ref{B}(ii), this then implies that $\lambda_{1,k} > 1$ for all $k$.
So, by Lemma \ref{B}(i), it follows that $\lambda_{1,k}$ is a strictly decreasing sequence and hence $\lambda_{1,k} \downarrow \lambda_1 > 1.$ 

Next, note that Lemma \ref{B} also implies that the sequences $\lambda_{1,k}$ and $\lambda_{\pdim,k}$ are both bounded away from $0$ and $\infty$. Hence, 
there exists a convergent subsequence $\bV_{k(j)} \to \bV \in \PDH(\pdim)$, with $\lambda_1(\bV) = \lambda_1 > 1$. Here, 
$\lambda_1(\bV) \ge \cdots \ge \lambda_\pdim(\bV) > 0$ denote the eigenvalues of $\bV$. Furthermore, by continuity, $\bV_{k(j)+1} \to G(\bV)$ with 
$\lambda_1\{G(\bV)\} = \lambda_1.$ However, Lemma \ref{B}(i) implies  $\lambda_1 = \lambda_1\{G(\bV)\} < \lambda_1(\bV) = \lambda_1$, a contradition.
Hence part (i) holds. The proof to part (ii) is analogous.
\endproof

By Lemma \ref{C} we have  $ 1 \le \liminf \lambda_{\pdim,k} \le \limsup \lambda_{1,k} \le 1$, which implies 
$\lim \lambda_{\pdim,k} = \lim \lambda_{1,k} = 1$. Thus,  $\bV_k \to \I_\pdim$.

\endproof

\section*{Proof of Theorem~\ref{th:oracle}}

\proof
Denote 
\beq \label{eq:tildeC}
\bo C = 
\frac{\pdim}{\ndim} \sum_{i=1}^\ndim \frac{\z_i \z_i^\hop}{\z_i^\hop \Mn^{-1} \z_i}   =  \Mn^{1/2} \Big( \frac{\pdim}{\ndim} \sum_{i=1}^\ndim \bo u_i \bo u_i^\hop \Big) \Mn^{1/2}
\eeq 
where $\bo u_i = \Mn^{-1/2}\z_i/\|\Mn^{-1/2}\z_i\|$ for $i=1,\ldots,\ndim$.  Hence the clairvoyant estimator is $\Mal= (1-\al)\bo C + \al \I$. 
First we note  that the MSE criterion is 
\begin{align*}
\Delta(\al) &= \E\big[ \|  \Mn^{-1} \Mal -  {\textstyle \frac{1}{\pdim} } \tr( \Mn^{-1} \Mal ) \I \|^2 \big]  \\ 
&=  \tr\Big( \Mn^{-2} \E\big[\Mal^2\big] \Big) - \frac 1 \pdim \E\big[ \tr^2(\Mn^{-1} \Mal ) \big]. 
\end{align*} 
Then observe that  
\begin{align*}
\tr(\Mn^{-1} \Mal) &= \tr\Big((1-\al)  \Mn^{-1/2} \Big( \frac{\pdim}{n} \sum_{i=1}^\ndim  \bo u_i \bo u_i^\hop \Big)  \Mn^{1/2} + \al \Mn^{-1} \Big)  \notag \\
&=\pdim (1-\al) + \al \tr(\Mn^{-1}) = \pdim  
\end{align*} 
where the 3rd identity   follows from the fact that  $\Tr(\Mn^{-1})=\pdim$. This result then implies that finding the minimum of  $\Delta(\al)$ is equivalent to finding the minimum of $\Delta^*(\al)= \tr\Big( \Mn^{-2} \E\big[\Mal^2\big] \Big)$.  

Next we show that a neat closed-form expression for $\Delta^*(\al)$ can be obtained by using the following identities: 
\begin{align}
&\E[\bo C] = \Mn  \label{eq:prop1} \\
&\E[ \bo C^2] = \frac{\pdim\{ \Mn^2 + \tr(\Mn) \Mn \} }{\ndim(\pdim+1)} + \Big(\frac{\ndim-1}{\ndim} \Big)\Mn^2 \label{eq:prop2} . 
\end{align}
The proofs rely on representation of $\bo C$ in \eqref{eq:tildeC} in terms of i.i.d. r.v.'s $\u_i$ which possess a uniform distribution on complex $\dim$-sphere and properties of their 
moments as stated in \cite[Lemma~4]{ollila_koivunen:2007e}.  Derivation is similar to the  Proof of Theorem~2 in \cite{chen_etal:2011} and is therefore omitted. 

Next note that 
\begin{align*}
\E[\Mal^2] &=  \E[((1-\al)\bo C + \al \I)^2] \\
& = 2 \al(1-\al) \E[\bo C] + \al^2 \I  + (1-\al)^2 \E[\bo C^2] 
\end{align*} 
and hence using \eqref{eq:prop1}, \eqref{eq:prop2} and the fact that $\tr(\Mn^{-1})=\pdim$, gives 
\begin{align*}
\Delta^*(\al)  =& 2 \al(1-\al) \pdim  + \al^2 \tr(\Mn^{-2}) \\
&{} \ + (1-\al)^2 \bigg\{ \frac{\pdim(\pdim + \pdim \tr(\Mn))}{\ndim(\pdim+1)}  + \Big(\frac{\ndim-1}{\ndim} \Big) \pdim \bigg\}  \\
 = & \al^2 ( \tr(\Mn^{-2}) - \pdim) + (1-\al)^2  \frac{\pdim(\pdim \tr(\Mn)-1)}{\ndim(\pdim+1)}  + C 
\end{align*} 
where a constant $C$ does not depend on $\al$. The minimizer $\al_o$ of $\Delta^*(\al)$ (and hence of $\Delta(\al)$) is thus  $\al_o = a/(a+b)$, where 
$a$ (resp. $b$) denotes the multiplier term of $(1-\al)^2$ (resp. $\al^2$) in the expression of $\Delta^*(\al)$  above. This then gives the stated result in the complex-valued case.

The proof for the real-case follows similarly, the only difference being that the idenitity in Eq. \eqref{eq:prop2}   in the real case is 
\[
\E[ \bo C^2] =  \frac{\pdim}{\ndim(\pdim+2)} \{ 2 \Mn^2 + \tr(\Mn) \Mn \} + \Big(\frac{\ndim-1}{\ndim} \Big)\Mn^2 . 
\]
\endproof



%\section{}

%\section*{Acknowledgment}

\ifCLASSOPTIONcaptionsoff
  \newpage
\fi

\end{document}